\documentclass{llncs}
\usepackage{times}
\usepackage{amsmath,gdefs,graph,alltt,ifthen}
\usepackage{graphics}
\usepackage{changebar}
\usepackage{latexsym, amssymb}

\newcommand{\Omit}[1]{}

\newboolean{TR}
\setboolean{TR}{true}
\ifthenelse{\boolean{TR}}{

\newcommand{\TrOnly}[1]{#1}
\newcommand{\SubOnly}[1]{}
\newcommand{\TrOnlyInFootnote}[1]{#1}
\newcommand{\TrOnlyInTable}[1]{#1}}
{

\newcommand{\TrOnly}[1]{}
\newcommand{\SubOnly}[1]{#1}
\newcommand{\TrOnlyInFootnote}[1]{}
\newcommand{\TrOnlyInTable}[1]{}}

\newcommand{\para}[1]{\vspace*{2ex}\noindent{\bf #1}\/}
\newcommand{\characteristicFormula}[1]{\xi^{#1}}
\newcommand{\NPcharacteristicFormula}[1]{\xi^{#1}}
\newcommand{\blur}{{\it canonical\/}}

\newcommand{\ty}{{\tt y\/}}
\newcommand{\ttt}{{\tt t}\/}
\newcommand{\te}{{\tt e\/}}

\newcommand{\tn}{{\tt n}}

\newcommand{\tList}{{\tt List}}
\newcommand{\tx}{{\tt x\/}}

\newcommand{\tinsert}{\texttt{insert\/}}

\newcommand{\figref}[1]{Fig.~\ref{Fi:#1}}
\newcommand{\tableref}[1]{Table~\ref{Ta:#1}}
\newcommand{\lemref}[1]{Lemma~\ref{Lem:#1}}
\newcommand{\defref}[1]{Definition~\ref{De:#1}}
\newcommand{\secref}[1]{Section~\ref{Se:#1}}
\newcommand{\equref}[1]{Eq.~(\ref{eq:#1})}
\newcommand{\exref}[1]{Example~\ref{Ex:#1}}
\newcommand{\theref}[1]{Theorem~\ref{The:#1}}
\newcommand{\appref}[1]{Appendix~\ref{Se:#1}}

\newcommand{\TRUE}{\boldsymbol{1}}
\newcommand{\FALSE}{\boldsymbol{0}}

\newcommand{\TC}[5]{(\mbox{{\it TC\/}}~{#1, #2}: #3)(#4, #5)}
\newcommand{\STRUCT}[1]{\mbox{3-STRUCT}[#1]}

\newcommand{\TSTRUCT}[1]{\mbox{2-STRUCT}[#1]}

\newcommand{\nodeFormula}[2]{\mbox{node}^{#1}_{{{#2}}}}

\newcommand{\Voc}{{\cal P}}
\newcommand{\PVar}{{{\it PVar\/}}}

\newcommand{\lsuperval}{\langle\!\langle}
\newcommand{\rsuperval}{\rangle\!\rangle}
\newcommand{\superval}[1]{\lsuperval{#1}\rsuperval}
\newcommand{\gammaHat}{\widehat{\gamma}}

\newcommand{\consistent}{admissible}
\newcommand{\ICA}{{\em ICA}}

\newcommand{\Lone}{EA(TC, f^1)}

\input{xy}
\xyoption{all}

\begin{document}
\title{Logical Characterizations of Heap Abstractions}

\author{G. Yorsh\inst{1}
 \and T. Reps\inst{2}
 \and M. Sagiv\inst{1}
 \and R. Wilhelm\inst{3}}
\institute{School of Comp.\ Sci.,
  Tel-Aviv University; \{gretay, msagiv\}@post.tau.ac.il
\and
  Comp.\ Sci.\ Dept., University of Wisconsin; reps@cs.wisc.edu
\and
  Informatik, Univ.\ des Saarlandes; wilhelm@cs.uni-sb.de
}

\maketitle

%\begin{spacing}{0.95}
\begin{abstract}
Shape analysis concerns the problem of determining ``shape invariants''
for programs that perform destructive updating on dynamically
allocated storage.  In recent work, we have shown how shape analysis
can be performed, using an abstract interpretation based on $3$-valued
first-order logic.  In that work, concrete stores are finite $2$-valued logical
structures, and the sets of stores that can possibly arise during
execution are represented (conservatively) using a certain family of
finite $3$-valued logical structures.
In this paper, we show how $3$-valued structures that arise
in shape analysis can be characterized using formulas in first-order logic
with transitive closure.
\TrOnly{We also define a non-standard (``supervaluational'') semantics for $3$-valued
first-order logic that is more precise than a conventional $3$-valued
semantics, and demonstrate that the supervaluational semantics can be effectively implemented
using existing theorem provers.}
\end{abstract}

%In this paper, we present results about
%the expressive power of $3$-valued logical structures.

\section{Introduction}

Abstraction and abstract interpretation \cite{kn:CC77}
are key tools for automatically verifying properties of systems,
both for hardware systems \cite{TOPLAS:CGL94,Thesis:Dams96}
and software systems \cite{Book:NNH99}.
In abstract interpretation, sets of concrete stores
are represented in a conservative manner by abstract values (as explained below).
Each transition of the system is given an interpretation
over abstract values that is conservative with respect to its
interpretation over corresponding sets of concrete stores;
that is, the result of ``executing'' a transition must
be an abstract value that describes a superset of
the concrete stores that actually arise.
This methodology guarantees that the results of abstract
interpretation overapproximate the sets of concrete stores that
actually arise at each point in the system.

One issue that arises when abstraction is employed
concerns the {\em expressiveness\/}
of the abstraction method:
``What collections of concrete states can be expressed exactly
using the given abstraction method?''
A second issue that arises when abstraction is employed
is how to {\em extract information\/} from an abstract value.
For instance, this is a fundamental problem for clients
of abstract interpretation, such as verification tools, program
optimizers, program-understanding tools, etc., which need to be
able to interpret what an abstract value means.
An abstract value $a$ represents a set of concrete stores $X$;
ideally, a query $\varphi$ should return an answer that summarizes the
result of posing $\varphi$ against each concrete store $S \in X$:
\begin{itemize}
  \item
    If $\varphi$ is true for each $S$, the summary answer should be ``true''.
  \item
    If $\varphi$ is false for each $S$, the summary answer should be ``false''.
  \item
    If $\varphi$ is true for some $S \in X$ but false for some $S' \in X$,
    the summary answer should be ``unknown''.
\end{itemize}

This paper presents results on both of these questions,
for a class of abstractions that originally arose in work on the
problem of shape analysis \cite{kn:JM81,kn:CWZ90,TOPLAS:SRW02}.
Shape analysis concerns the problem of finding ``shape descriptors''
that characterize the shapes of the data structures that a program's
pointer variables point to.
Shape analysis is one of the most challenging problems in abstract
interpretation because it generally deals with programs written
in languages like C, C++, and Java, which allow
(i)~dynamic allocation and deallocation of cells from the heap,
(ii)~destructive updating of structure fields, and, in the case of Java,
(iii)~dynamic creation and destruction of threads.
This combination of features creates considerable difficulties for any
abstract-interpretation method.

The motivation for the present paper was to understand the
expressiveness
of the shape abstractions defined in \cite{TOPLAS:SRW02}.
%The motivation for the present paper was to understand the
%expressiveness
%of the shape abstractions defined in \cite{TOPLAS:SRW02}.
In that work, concrete stores are finite $2$-valued logical structures, and
the sets of stores that can possibly arise during execution are
represented (conservatively) using a certain family of finite $3$-valued
logical structures.
In this setting, an abstract value is a set of $3$-valued logical structures.
Because the notion of abstraction used in \cite{TOPLAS:SRW02}
is based on logical structures, our results are actually much more broadly
applicable than shape-analysis problems.
For example, in~\cite{Yahav:2001:VSP}) is applicable to accurately
model concurrency in Java programs which contain dynamic creation of objects and
threads.
In fact our results apply to
any abstraction in which concrete states of a system are represented by
finite $2$-value logical structure and abstraction is performed
via the mechanisms described in Sections~\ref{Se:Preliminaries}
and~\ref{Se:Classic}. Throughout the paper, however,
we use shape-analysis examples to illustrate the concepts
discussed.

The paper investigates the
expressiveness of finite
$3$-valued structures by giving a logical characterization of these structures;
that is, we examine the question
\begin{quote}
  For a given $3$-valued structure $S$,under what
  circumstances is it possible to create a formula $\gammaHat(S)$,
  such that $\C{S}$ satisfies $\gammaHat(S)$ exactly when
  $\C{S}$ is a $2$-valued structure that $S$ represents?  I.e.,
  $\C{S} \models \gammaHat(S)~~\text{iff}~~S~\text{represents}~\C{S}$.
\end{quote}
This paper presents two results concerning this question:
%Although it is not possible to give a formula $\gammaHat(S)$ for an arbitrary structure $S$,
%if $\gammaHat(S)$ is to be written in first-order logic with transitive closure,
%this paper presents two results concerning this question:
\begin{itemize}
  \item
    It is not possible to give a formula $\gammaHat(S)$ written in first-order logic with transitive closure
    for an arbitrary structure $S$.
    However,  it is always possible for a well-defined class of $3$-valued structures.
    (This class includes all the $3$-valued structures
    that have been shown to
    be useful for shape analysis \cite{TOPLAS:SRW02}.)
  \item
    Moreover, it is always possible to give a $\gammaHat(S)$ in general,
    using a more powerful formalism, namely, monadic
    second-order formulas.
\end{itemize}

The ability to write a formula $\gammaHat(S)$ that exactly captures what
$S$ represents provides a fundamental tool
for improving TVLA \cite{SAS:LS00} by the use of symbolic methods.
The current TVLA system performs iterative fixed-point computations and
yields at every program point a set of $3$-valued structures,
which represent a superset of all possible stores that can arise at this point in any execution.
However, TVLA suffers from two limitations:
(i)~it is not always as precise as possible (as explained below);
(ii)~it does not scale to handle large programs, because the worst-case complexity of the
algorithm is doubly-exponential in
certain parameters (typically, the number of program variables).

The contributions of this paper lay the required groundwork
for using symbolic techniques to address both of these limitations.
The ability to characterize a $3$-valued structure $S$ by a formula $\gammaHat(S)$
is a key step toward harnessing a standard ($2$-valued) theorem prover
to aid in abstract interpretation:
%In particular, it permits the precision and scalability problems to be addressed by
\begin{itemize}
  \item
    Computing the effect of a program statement on an abstract value in the
    most-precise way possible %(relative to the abstraction in use)
    for a given shape-analysis abstraction.
  \item
    Developing a modular shape-analysis by using
    \emph{assume-guarantee} reasoning.
    The idea is to allow arbitrary first-order formulas
    to be used to express pre- and post-conditions,
    thereby enabling the code of each procedure to be analyzed once
    for all potential contexts.
    This allows to scale shape analysis and to apply to applications in which not
    all the source code is available. This becomes specifically
    profitable for recursive procedures since it saves the need to iterate
    shape analysis.
\end{itemize}
These methods are the subject of \cite{TACAS:YRS04,Hob:cc05}.

Another contribution of this paper directly addresses
the first of the aforementioned limitations of TVLA's current technique.
We give a procedure for extracting information from a $3$-valued logical structure $S$
in the most-precise way possible.
That is, we give a nonstandard way to check if a formula $\varphi$
holds in $S$:
    \begin{itemize}
      \item
        If $\gammaHat(S) \implies \varphi$ is valid, i.e., holds in all
        $2$-valued structures, we know that $\varphi$
        evaluates to $1$ in all the $2$-valued structures represented by
        $S$.
      \item
        If $\gammaHat(S) \implies \neg \varphi$ is valid, we know that $\varphi$
        evaluates to $0$ in all the $2$-valued structures represented by
        $S$.
      \item
        Otherwise we know that there exists a $2$-valued structure represented by $S$
        where $\varphi$ evaluates to $1$, and there exists another $2$-valued structure
        represented by $S$ where $\varphi$ evaluates to $0$.
    \end{itemize}
This method represents the most-precise way of extracting information
from a $3$-valued logical structure; in particular, whenever this
method returns $1/2$ (standing for ``unknown''),
any sound method for extracting information
from $S$ must also return $1/2$. This is in contrast
with the techniques used in \cite{TOPLAS:SRW02},
which can return $1/2$ even when all the $2$-valued structures
represented by $S$ have the value $1$ (or all have the value $0$).

Although the validity question is undecidable for
first-order logic with transitive closure,
several theorem provers for first-order logic have been created.
We report on two experiments in which we used these tools to implement
symbolic procedures for extracting information from a $3$-valued
structure in the most-precise way possible.
Also, in~\cite{csl04:IRRSY}, we have identified a decidable subset of first-order
logic with transitive closure that is useful for shape analysis.
We define conditions under which $\gammaHat$ can be expressed in that logic.

The remainder of the paper is organized as follows.
\secref{Preliminaries} defines our terminology,
and explains the use of $3$-valued structures
as abstractions of $2$-valued structures.
\secref{Classic}
presents the results on the expressiveness of $3$-valued structures,
and gives an algorithm for generating $\gammaHat$ for certain families of
3-valued structures.
\secref{Supervaluation} discusses the problem of reading out
information from a $3$-valued structure in the most-precise way possible.
\secref{Applications} discusses the applications of $\gammaHat$
to program analysis and some implementation issues.
\secref{RelatedWork} discusses related work.
%describes an experiment in which we used
%the $\gammaHat$ operation and an existing theorem prover
%for first-order logic to read out information from
%a $3$-valued structure.
\appref{Canonic} defines an alternative abstract domain for shape analysis,
based on canonical abstraction, and the $\gammaHat$
operation for that domain.
\appref{NPFormula} shows how to characterize general $3$-valued structures.
\appref{AppendixTable} contains the details for one of the paper's examples.
%\secref{FinalRemarks} makes some final remarks.
%\appref{AppendixTable} demonstrates how to use the $3$-valued structures
%obtained from a TVLA analysis to construct a loop invariant;
%this is then used to show that certain properties of a linked data
%structure hold on each loop iteration.
The proofs appear in \appref{Proofs}.
   %introduction
\section{Preliminaries}\label{Se:Preliminaries}

\secref{Syntax} defines the syntax and standard Tarskian semantics
of first-order logic with transitive closure and equality.
\secref{Integrity} introduces \emph{integrity formulas},
which exclude structures that do not represent a potential store.
\secref{Semantics} introduces $3$-valued logical structures,
which extend ordinary logical structures with an extra value, $1/2$, which
represents ``unknown'' values that arise when several concrete nodes are
represented by a single abstract node.
The powerset of $3$-valued structures forms an abstract domain, which is related to the concrete
domain consisting of the powerset of $2$-valued structures via \textit{embedding}, as described in \secref{Embedding}.

\figref{Insert}(a) shows the declaration of a linked-list
data type in C, and \figref{Insert}(b) shows
a C program that searches a list and splices a new element into the list.
This program will be used as a running example throughout this paper.
\begin{figure}
\resizebox{\textwidth}{!}{
\framebox{
\begin{tabular}{@{\hspace{1ex}}c@{\hspace{6ex}}c@{\hspace{1ex}}}
\begin{minipage}{1in}
\begin{alltt}
\begin{tabbing}
/* list.h */ \\
ty\=pedef struct node \{ \+ \\
        struct node *n;  \\
        int data;      \- \\
\} *List;
\end{tabbing}
\end{alltt}
\end{minipage}
&
\begin{minipage}{1in}
\begin{alltt}
\begin{tabbing}
/* insert.c */ \\
\#include "list.h" \\
vo\=id insert(List x, int d) \{ \+ \\
     List y, t, e; \\
     as\=sert(acyclic\_list(x) \&\& x != NULL);\\
     y = x; \\
     while (y->n != NULL \&\& ...) \+\\
       y = y->n; \-\\
     t = malloc();\\
     t->data = d;\\
     e = y->n;\\
     t->n = e;\\
     y->n = t; \-\\
\}
\end{tabbing}
\end{alltt}
\end{minipage}
\\
(a) & (b)\\
\end{tabular}}}
\caption{\label{Fi:Insert}(a)~Declaration of a linked-list data type in C.
(b)~A C function that searches a list pointed to by parameter {\tt x},
and splices in a new element.}
\end{figure}

\subsection{Syntax and Semantics of First-Order Formulas with Transitive Closure}
\label{Se:Syntax}

%For us, concrete stores are {\em logical structures}.
We represent concrete stores by ordinary $2$-valued logical structures
over a fixed finite set of predicate symbols
$\Voc = \{eq, p_1, \ldots, p_n \}$, where $eq$ is a designated binary predicate,
denoting equality of nodes.
We also use $maxR$ to denote the maximal arity of the predicates in $\Voc$.
Without loss of generality we exclude
constant and function symbols from the logic.\footnote{Constant
symbols can be encoded via unary predicates, and $n$-ary functions
via $(n+1)$-ary predicates.}

\begin{example}
\tableref{Predicates} lists the set of predicates used in the running
example.
%%pointed-to-by
The unary predicates $x$, $y$, $t$, and $e$
correspond to the program variables \tx, \ty, \ttt, and \te,
respectively.
%% n
The binary predicate $n$ corresponds to the \tn\ fields of {\tt List\/}
elements.
%% is
The unary predicate $is$ (``is shared'') captures ``heap sharing'',
i.e., {\tt List\/} elements pointed to by more than one field.
(It was introduced in~\cite{kn:CWZ90}
%and also used in~\cite{kn:SRW98}
to capture list and tree data structures.)
% reach
The unary predicates $r_x$,
$r_y$, $r_t$, and $r_e$ hold for heap nodes reachable from the program
variables \tx, \ty, \ttt, and \te, respectively.
A heap node $u$ is said to be {\em reachable\/}
from a program variable if the variable points to a heap node $u'$,
and it is possible to go from $u'$ to $u$ by following
zero or more \tn-links.
Reachability is defined in term of  the reflexive
transitive closure of the predicate $n$.

The notion of reachability plays a crucial role in defining abstractions that are useful
for proving program properties in practice.
For instance, it may have the effect of preventing disjoint lists from being collapsed
in the abstract representation.
This may significantly
improve the precision of the answers obtained by a program analysis.
%The predicates $n$ and $is$ are used to model
%other programs that use the {\tt List\/}
%data-type declaration from \figref{Insert}(a).

\begin{table}
\begin{center}
\begin{tabular}{|@{\hspace{1ex}}l@{\hspace{1ex}}|@{\hspace{1ex}}l@{\hspace{1ex}}|}
\hline
{\bf Predicate\/} & {\bf Intended Meaning\/}\\
\hline
  $eq(v_1, v_2)$   & Do $v_1$ and $v_2$ denote the same heap node? \\
  \hline
  $q(v)$   & Does pointer variable {\tt q} point to node $v$? \\
  \hline
  $n(v_1, v_2)$   & Does the {\tt n} field of $v_1$ point to $v_2$? \\
  \hline
  $is(v)$ & Is $v$ pointed to by more than one field ?\\
  \hline
 % $c(v)$ & Is $v$ located on a cycle of field pointers ?\\
 % \hline
  $r_q(v)$ & Is the node $v$ reachable from {\tt q} ?\\
\hline
\end{tabular}
\end{center}
\caption{\label{Ta:Predicates}The set of predicates for representing the stores
manipulated by programs that use the {\tt List\/} data-type
from \figref{Insert}(a).
$q$ denotes an arbitrary predicate in the set $PVar$,
which contains a predicate for each
program variable of type {\tt List\/}.
In the case of \tinsert, $\PVar = \{{\tt x, y, t, e}\}$.}
\end{table}
\end{example}

We define first-order formulas inductively over the {\bf
vocabulary\/} $\Voc$ using the logical connectives $\lor$ and $\neg$,
the quantifier $\exists$, and the operator `${\it TC}$' in the standard way:
\begin{eqnarray*}
%\begin{array}{l@{\hspace{1.2cm}}l@{\hspace{1.2cm}}rcl}
\varphi  ::= \FALSE \mid \TRUE \mid p(v_1, \ldots, v_k) \mid (\neg
\varphi_1) \mid (\varphi_1 \lor \varphi_2) %\nonumber\\
\mid (\exists v_1: \varphi_1) \mid
\TC{v_1}{v_2}{\varphi_1}{v_3}{v_4} \\
where~p \in \Voc;  v_i \mbox{~are variables};
\varphi, \varphi_i
\mbox{~are formulas} %\nonumber
%\end{array}
\end{eqnarray*}

The set of free variables of a formula is defined as usual.
A formula is {\bf closed\/} when it has no free variables.
The operator `${\it TC}$' denotes transitive closure.
If $\varphi_1$ is a formula with free variables $V$, then
$\TC{v_1}{v_2}{\varphi_1}{v_3}{v_4}$ is a formula with free
variables $(V - \{v_1, v_2\}) \cup \{v_3, v_4\}$.

We use several shorthand notations: $\varphi_1 \implies \varphi_2
\eqdef (\neg \varphi_1 \lor \varphi_2)$; $\varphi_1 \land
\varphi_2 \eqdef \neg (\neg \varphi_1 \lor \neg \varphi_2)$;
$\varphi_1 \impliesBothWays \varphi_2 \eqdef (\varphi_1 \implies
\varphi_2) \land (\varphi_2 \implies \varphi_1)$; and $\forall v:
\varphi \eqdef \neg \exists v: \neg \varphi$.
The transitive closure of a binary predicate $p$ is $p^+(v_3, v_4) \eqdef \TC{v_1}{v_2}{p(v_1, v_2)}{v_3}{v_4}$.
The {\em reflexive } transitive closure of a binary predicate $p$ is $p^*(v_3, v_4)
\eqdef (\TC{v_1}{v_2}{p(v_1, v_2)}{v_3}{v_4}) \lor eq(v_3, v_4)$.
The order of precedence among the connectives, from highest to
lowest, is as follows: $\neg$, $\land$, $\lor$, `${\it TC}$',
$\forall$, and $\exists$.
We drop parentheses wherever possible, except for emphasis.

\begin{definition}\begin{Name}$2$-valued Logical Structures\end{Name}
\label{De:Logical}
Let $\Voc_i$ denote the set of predicate symbols with arity $i$.
A {\bf logical structure over $\Voc$\/} is
a pair $S = \B{U, \iota}$ in which
\begin{itemize}
  \item
    $U$ is a (possibly infinite) set of nodes.
  \item
    $\iota$ is the interpretation of predicate symbols, i.e., for
    every predicate symbol $p \in \Voc_i$,
    $\iota(p)\colon U^i \to \{0, 1\}$ determines the tuples for which $p$ holds.
    Also, $\iota(eq)$ is the interpretation of equality,
    i.e., $\iota(eq)(u_1, u_2) =1$ iff $u_1 = u_2$.
\end{itemize}
%We denote the (infinite) set of structures by $\ConcreteStruct{\Voc}$.
\end{definition}

Below we define the standard Tarskian semantics for first-order logic.
\begin{definition}\begin{Name}Semantics of First-Order Logical Formulas\end{Name}
Consider a logical structure $S = \B{U, \iota}$.
An {\bf assignment\/} $Z$ is a function that maps free variables to
nodes (i.e., an assignment has the functionality
$Z\colon \{v_1, v_2, \ldots \} \to U$).
An assignment that is defined on all free
variables of a formula $\varphi$ is called {\bf complete} for
$\varphi$.  In the sequel, we assume that every assignment $Z$ that arises
in connection with the discussion of some formula $\varphi$ is complete
for $\varphi$.
We say that $S$ and $Z$ {\bf satisfy\/} a formula $\varphi$ (denoted by $S, Z \models \varphi$)
when one of the following holds:
\begin{itemize}
  \item
    $\varphi \equiv \TRUE$
  \item
    $\varphi \equiv p(v_1, v_2, \ldots, v_i)$ and $\iota(p)(Z(v_1), Z(v_2), \ldots, Z(v_i)) = 1$.
  \item
    $\varphi \equiv \neg\varphi_0$ and $S, Z \models \varphi_0$ does not hold.
  \item
    $\varphi \equiv \varphi_1 \lor \varphi_2$, and either
    $S, Z \models \varphi_1$ or $S, Z \models \varphi_2$.
  \item
    $\varphi \equiv \exists v_1: \varphi_1$ and there exists a node
    $u \in U$, $m \geq 2$, such that $S, Z[v_1 \mapsto u] \models \varphi_1$.
  \item $\varphi \equiv \TC{v_1}{v_2}{\varphi_1}{v_3}{v_4}$
  and there exists $u_1, u_2, \ldots, u_m \in U$, $m \geq 2$, such that
  $Z(v_3)=u_1$, $Z(v_4)=u_m$ and for all $1 \leq i < m$,
  $S, Z[v_1 \mapsto u_i, v_2 \mapsto u_{i+1}] \models \varphi_1$.
\end{itemize}

For a closed formula $\varphi$, we will omit the assignment in
the satisfaction relation, and merely write $S \models \varphi$.
%We also use the notation $\semp{\varphi}$ to denote
%the set of concrete structures that satisfy $\varphi$:
%$\semp{\varphi} = \{ S \mid S \in \ConcreteStruct{\Voc}, S \models \varphi\}$.
\end{definition}
%{
%We use standard Tarskian semantics for first-order logic:
%Let $S = \B{U, \iota}$ be a first order structure.
%Let $\varphi$ be a formula and $Z$ be an assignment
%from the free variables of $\varphi$ into $U$.
%We say that $S$ and $Z$ {\bf satisfy\/} $\varphi$
%(denoted by $S, Z \models \varphi$)
%when $\varphi$ evaluates to $1$ on $S$ (with the
%free variables of $\varphi$ bound to the nodes
%specified by $Z$).
%For a closed formula $\varphi$, we omit the assignment in
%the satisfaction relation, and merely write $S \models \varphi$.
%}

\subsection{Integrity Formula}
\label{Se:Integrity}

Because not all logical structures represent stores, we use a
designated closed formula $F$, called the \emph{integrity
formula},\footnote{In \cite{TOPLAS:SRW02} these are called ``hygiene
conditions''.} to exclude structures that are not of interest; in our application, such
structures are ones that do not correspond to possible stores.
This allows us to restrict the set of structures to the ones satisfying $F$.

\begin{definition}
A structure $S$ is {\bf \consistent} if $S \models F$.
\end{definition}

In the rest of the paper, we assume that we work with a fixed integrity formula $F$.
All our notations are parameterized by $\Voc$ and $F$.
\begin{example}\label{Ex:IntegrityExample}
For the $\tList$ data type, there are four conditions that define the admissible structures.
At any time during execution,
\begin{description}
\item [{\rm (a)}] each program variable can point to at most one heap node.
\item [{\rm (b)}] the {\tt n} field of a heap node can point to at most one heap node.
\item [{\rm (c)}] predicate $is$ (``is shared'') holds for exactly
those nodes that have two or more predecessors.
%\item [{\rm (d)}] Predicate $\pc$ (``cyclic'') holds for those nodes that are located on a cycle.
\item [{\rm (d)}] the reachability predicate for each variable {\tt q} holds for exactly those nodes
that are reachable from program variable {\tt q}.
\end{description}
The set $\PVar$ contains a predicate for each
program variable of type {\tt List\/}; in the case of \tinsert, $\PVar = \{{\tt x, y, t, e}\}$.
Thus, the integrity formula $F_{List}$ for the {\tt List} data-type is:
\[
\begin{array}{@{\hspace{0ex}}l@{\hspace{1.0ex}}r@{\hspace{2.0ex}}l@{\hspace{0ex}}}
&\land_{p \in \PVar} \forall v_1, v_2: p(v_1) \land p(v_2)
\implies eq(v_1, v_2) & (a)\\
\land & \forall v, v_1, v_2: n(v, v_1)
\land n(v, v_2)
\implies eq(v_1, v_2) & (b) \\
\land & \forall v: is(v) \iff
\exists v_1, v_2: \neg eq(v_1, v_2)
\land n(v_1, v) \land n(v_2, v) & (c)\\
%\left ( \exists v_1, v_2:
%\begin{array}{ll}
%& \neg eq(v_1, v_2)\\
%\land & n(v_1, v)
%\land n(v_2, v)
%\end{array}
%\right ) & (d)\\
%\land & \forall v: c(v) \iff n^+(v,v) & (4) \\
\land & \land_{q \in \PVar} \forall v: r_q(v) \iff \exists v_1:
q(v_1) \land n^*(v_1, v) & (d)\\
\end{array}
\]
\end{example}

\subsection{$3$-Valued Logical Structures and Embedding} %Canonical
\label{Se:Semantics}

In this section, we define $3$-valued logical structures,
which provide a way to represent a set of $2$-valued logical structures
in a compact and conservative way.

We say that the values $0$ and $1$ are {\em definite values\/} and
that $1/2$ is an {\em indefinite value\/}, and define a partial order
$\sqsubseteq$ on truth values to reflect information content.
$l_1 \sqsubseteq l_2$ denotes that $l_1$ possibly has more definite
information than $l_2$:
\begin{definition}{\bf [Information Order]}.\label{De:ThreeValuedOrder}
  For $l_1, l_2 \in \{0, 1/2, 1\}$, we define the {\bf information
  order\/} on truth values as follows:
  $l_1 \sqsubseteq l_2$ if $l_1 = l_2$ or $l_2 = 1/2$.
  %The symbol $\sqcup$ denotes the least-upper-bound operation
  %with respect to $\sqsubseteq$.
\end{definition}
\begin{definition}
A {\bf 3-valued logical structure\/} over
$\Voc$ is the generalization of $2$-valued
structures given in \defref{Logical}, in that predicates may have
the value $1/2$.
This means that $S= \B{U, \iota}$ where for $p \in \Voc_i$,
$\iota(p)\colon ({U^S})^i \to \{ 0, 1, 1/2 \}$.
In addition, (i)~for all $u \in U^S$, $\iota^S(eq)(u, u)
\sqsupseteq 1$, and (ii)~for all $u_1, u_2 \in U^S$ such that
$u_1$ and $u_2$ are distinct nodes,
$\iota^S(eq)(u_1, u_2) = 0$.

A node $u \in U$ having $\iota^S(eq)(u, u)=1/2$ is called a {\bf summary node\/}.
As we shall see, such a node may represent more
than one node from a given $2$-valued structure.
\end{definition}

We denote the set of $2$-valued logical structures
%that represent valid stores
by $\TSTRUCT{\Voc}$.
The set of 3-valued logical structures is denoted by $\STRUCT{\Voc}$.
%Note that this refines the notations $\STRUCT{\Voc}$ and
%$\TSTRUCT{\Voc}$ used in~\cite{TOPLAS:SRW02}, limiting the sets to
%include only \emph{\consistent} structures.

A $3$-valued structure can be depicted as a directed graph, with
nodes as graph nodes.
A unary predicate $p$ is represented in the graph by having
a solid arrow from the predicate name $p$ to node $u$ for each
node $u$ for which $\iota(p)(u) = 1$.
An arrow between two nodes indicates whether a binary predicate holds
for the corresponding pair of nodes.
An indefinite value of a predicate is shown by a dotted arrow;
the value $1$ is shown by a solid arrow;
and the value $0$ is shown by the absence of an arrow.

\begin{example}\label{Ex:threeStructEx}
\figref{threeStruct}(d) shows a $3$-valued structure that
represents possible inputs of the \tinsert\/ program.
This structure represents all lists
that are pointed to by program variable \tx\ and
have at least two elements.
The structure has $2$ nodes, $u_1$ and $u_2$,
where $u_1$ is the head of the list pointed to by \tx,
and $u_2$ is a summary node (drawn as a double circle),
which represents the tail of the list.
Predicate $r_x$ holds for $u_1$ and $u_2$, indicating that
all elements of the list are reachable from \tx.
Other unary predicates are not shown, indicating that their values are $0$
for all nodes, i.e., the program variables  \ty, \te, and \ttt\/ are {\tt NULL},
and there is no sharing in the list.
The dotted edge from $u_1$ to $u_2$ indicates that there may be \tn\/-links from
the head of the list to some elements in the tail.
In fact, the $(u_1,u_2)$-edge represents exactly one \tn\/-link that points to exactly one
list element, because of conjunct (b) of the integrity formula \exref{IntegrityExample}.
In contrast, the dotted self-loop on $u_2$
represents all \tn\/-links that may occur in the tail.

%For example, the formula:
%\begin{equation}
%\exists v_1, v_2: x(v_1) \land n(v_1, v_2)
%\label{eq:NextX}
%\end{equation}
%evaluates to $1/2$ on this structure. The next section will
%present a more precise way to obtain the value $1$ for this
%formula, which is in line with the fact that in this structure
%\texttt{x->n} must be allocated.
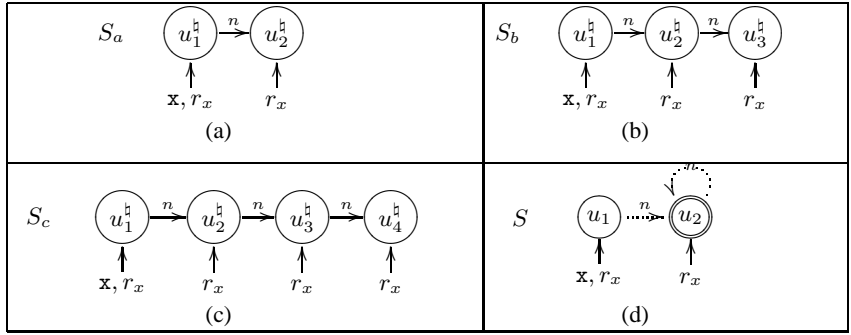
\begin{figure}
\begin{center}
%\framebox{
\begin{tabular}{|@{\hspace{1ex}}c@{\hspace{6ex}}|c@{\hspace{6ex}}|}
\hline
\xymatrix@R9pt@C10pt{
   %\xyEdge &
   S_a
   &
     \xyNonSummaryNode{\Conc{u}_1}
             %\xyDottedLabeledEdge{n}
             \ar[r]^{n}
    &
           \xyNonSummaryNode{\Conc{u}_2}
            %\ar@{.>}@(ur, ul)[]|{n}
            \\
  &    \tx, r_{x}\ar[u] &
            r_{x}\ar[u] &
}
&
\xymatrix@R9pt@C10pt{
   %\xyEdge &
   S_b
   &
     \xyNonSummaryNode{\Conc{u}_1}
             %\xyDottedLabeledEdge{n}
             \ar[r]^{n}
    &
           \xyNonSummaryNode{\Conc{u}_2}
%           \ar@{.>}@(ur, ul)[]|{n} \\
             %\xyDottedLabeledEdge{n}
             \ar@{>}[r]^{n}
      &
           \xyNonSummaryNode{\Conc{u}_3} \\
%             %\xyDottedLabeledEdge{n}
%             \ar@{.>}[r]^{n}
%    &
%           \xySummaryNode{u_4}
%           %\xyDottedLabeledSelfEdge{n}
%           \ar@{.>}@(ur, ul)[]|{n}\\
           &        \tx, r_{x}\ar[u] &
                    r_{x}\ar[u] &
                    r_{x}\ar[u]
%                   \mbox{$\begin{array}{c}\ty,
%                             r_{x},\\ r_{y}\end{array}$}\ar[u]
%                   &  r_{x}, r_{y}\ar[u]
}
\\
(a) & (b)\\[6pt]
\hline
\xymatrix@R9pt@C12pt{
   S_c
   &
     \xyNonSummaryNode{\Conc{u}_1}
             \ar[r]^{n}
    &
           \xyNonSummaryNode{\Conc{u}_2}
             \ar@{>}[r]^{n}
      &
           \xyNonSummaryNode{\Conc{u}_3}
             \ar@{>}[r]^{n}
      &
           \xyNonSummaryNode{\Conc{u}_4} \\
           &        \tx, r_{x}\ar[u] &
                    r_{x}\ar[u] &
                    r_{x}\ar[u] &
                    r_{x}\ar[u]
}
&
\xymatrix@R9pt@C12pt{
   %\xyEdge &
   S
   &
     \xyNonSummaryNode{u_1}
             %\xyDottedLabeledEdge{n}
             \ar@{.>}[r]^{n}
    &
           \xySummaryNode{u_2} \ar@{.>}@(ur, ul)[]|{n} \\
   &             \tx, r_{x}\ar[u] &
                    r_{x}\ar[u] &
}
\\
(c) & (d) \\
\hline
\end{tabular}
%}
\end{center}
\caption{\label{Fi:threeStruct} (a),(b),(c)~Examples
of $2$-valued structures representing linked-lists that are
pointed to by program variable \tx,
of length $2$, $3$, and $4$, respectively.
(d)~$S$ represents all lists that are pointed to
by program variable \tx\ and that have at
least two elements, including the lists represented by (a)-(c).
%In fact, $S$ represents possible inputs of the \tinsert\/ program.
}
\end{figure}
\end{example}

%\subsection{Embedding into $3$-Valued Structures}
%\label{Se:Embedding}
%
%In this section, we introduce the concept of {\em embedding}, which
%provides a way to relate $2$-valued and $3$-valued structures,
%and formulate the Embedding Theorem, which
%relates $2$-valued and $3$-valued interpretations of a given formula.

\subsection{Embedding Order}\label{Se:Embedding}

We define the {\em embedding ordering\/} on structures as follows:
\begin{definition}\label{De:Embedding}
Let $S = \B{U^S, \iota^S}$ and ${S'} = \B{U^{S'}, \iota^{S'}}$ be
two logical structures, and let $f\colon U^S \to U^{S'}$ be a surjective.
We say that $f$ {\bf embeds\/} $S$ in ${S'}$ (denoted by
$S \sqsubseteq^{f} {S'}$) if for every predicate symbol
$p \in \Voc_i$ and all $u_1, \ldots, u_i \in U^S$,
\begin{equation}
  \label{eq:EmbeddingCondition}
  \iota^S(p)(u_1, \ldots, u_i)
  \sqsubseteq
  \iota^{S'}(p)(f(u_1), \ldots, f(u_i))
\end{equation}

We say that $S$ {\bf can be embedded in\/} $S'$ (denoted by
$S \sqsubseteq S'$) if there exists a function $f$ such that
$S \sqsubseteq^f S'$.
\end{definition}

\begin{example}\label{Ex:threeStructEmbed}
\figref{threeStruct}(a)-(c) show some of the $2$-valued structures that
can be embedded into the $3$-valued structure $S$ shown in \figref{threeStruct}(d).
The function that embeds $S_a$ into $S$ maps the node $\Conc{u}_i \in U^{S_a}$ to $u_i \in U^{S}$, for
$i=1,2$.
The function that embeds $S_b$ into $S$ maps the node $\Conc{u}_1 \in U^{S_b}$ to $u_1 \in U^{S}$,
and both $\Conc{u_2}, \Conc{u_3} \in U^{S_b}$ to $u_2 \in U^{S}$.
Also, \equref{EmbeddingCondition} holds, because whenever a predicate
has a definite value in $S$, the corresponding predicate in $S_b$
has the same value.
For example, $\iota^S(x)(u_2)$ is $0$ and $f(\Conc{u_2})=f(\Conc{u_3})=u_2$,
and both $\iota^{S_b}(x)(\Conc{u_2})$ and $\iota^{S_b}(x)(\Conc{u_3})$ are $0$.
Similarly, $\iota^S(r_x)(u_2) = 1$, and both
$\iota^{S_b}(r_x)(\Conc{u_2})$ and $\iota^{S_b}(r_x)(\Conc{u_3})$ are $1$.
For a binary predicate, $\iota^S(n)(u_2, u_1) = 0$, and both
$\iota^{S_b}(n)(\Conc{u_2}, \Conc{u_1})$ and $\iota^{S_b}(n)(\Conc{u_3}, \Conc{u_1})$ are $0$.
\end{example}

\begin{Remark}
Embedding can be viewed as a variant of homomorphism~\cite{homomorphism}.
In cases where $S$ is a $2$-valued structure
(i.e., all predicates in $S$ have definite values, including $eq$, which is interpreted as
standard equality), checking whether a $2$-valued structure $S'$ embeds
into $S$ is equivalent to checking whether there is an isomorphism between $S'$ and $S$.
In cases where all nodes in $S$ are summary nodes (i.e., for all $u \in U^S$,
$\iota^S(eq)(u,u) = 1/2$), and all other values of predicates are definite,
embedding is equivalent to strong homomorphism.
In cases where all nodes in $S$ are summary nodes and all other values
of predicates are either $0$ or $1/2$, embedding is equivalent to homomorphism.
In all other cases, i.e, when a predicate value for some tuple in $S$ is 1,
embedding generalizes the notion of homomorphism.
\end{Remark}

\begin{Remark}
In \defref{Embedding}, we require that $f$ be surjective in order to
guarantee that a quantified formula, such as $\exists v: \varphi$,
has consistent values in two $3$-valued structures $S$ and $S'$ related by embedding.
For example, if $f$ were not surjective,
then there could exist an individual $u' \in U^{S'}$, not in the range
of $f$, such that the value of $S'$ on $\varphi$ is $1$ when $v$ is assigned to $u'$.
This would permit there to be structures $S$ and $S'$ for which
the value of $\exists v: \varphi$ on $S$ is $0$ but
its value on $S'$ is $1$.
\end{Remark}

\para{Concretization of $3$-Valued Structures.}
Embedding allows us to define the (potentially infinite)
set of concrete structures that a set of $3$-valued structures
represents:
\begin{definition}\label{De:Concrete}
\begin{Name}Concretization of $3$-Valued Structures\end{Name}
For a set of structures $X \subseteq  \STRUCT{\Voc}$,
we denote by $\gamma(X)$ the set of $2$-valued structures that $X$ represents,
i.e.,
\begin{equation} \label{eq:Concretization}
\begin{array}{r}
  \gamma(X) = \{ \C{S} \in \TSTRUCT{\Voc} \mid
  \text{exists}~ S \in X
  \text{such that}~\C{S} \sqsubseteq S~\text{and}~\C{S} \models F\}
\end{array}
\end{equation}

Also, for a singleton set $X = \{S\}$ we write $\gamma(S)$ instead of $\gamma(X)$.
\end{definition}
\begin{example}
\exref{threeStructEmbed} shows that $S_a \sqsubseteq S$, $S_b \sqsubseteq S$,
and $S_c \sqsubseteq S$ for the $2$-valued structures
in Figs.~\ref{Fi:threeStruct}(a-c); also, the integrity formula is
satisfied for $S_a$, $S_b$, and $S_c$.
Therefore, $S_a$, $S_b$, and $S_c$ are in the concretization of $3$-valued structure $S$:
$S_a, S_b, S_c \in \gamma(S)$.
Note that the indefinite values of predicates in $S$
allow the corresponding values in $S_b$ to be either $0$ or $1$.
In particular, $\iota^{S}(eq)(u_2, u_2) = 1/2$ reflects the
fact that the abstract node $u_2$ may represent more than
one concrete node.
Indeed, $S_b$ contains two nodes, $\Conc{u_2}$ and $\Conc{u_3}$,
that are represented by $u_2 \in S$. Also,
$\iota^{S}(eq)(\Conc{u_2}, \Conc{u_3}) = 0$,
but $\iota^{S}(eq)(\Conc{u_2}, \Conc{u_2}) = 1$.
\end{example}
The abstract domain we consider is the powerset of 3-valued structures, where the ordering relation $\sqsubseteq$ is defined as follows:
for every two sets of $3$-valued structures $X_1$ and $X_2$,
$X_1 \sqsubseteq X_2$ iff for all $S_1 \in X_1$  there exists $S_2 \in X_2$ such that $S_1$ is embedded into $S_2$.
%Hoare order based on embedding.

\subsubsection{The Analysis Technique}
The TVLA (\cite{SAS:LS00}) system carries out an abstract interpretation~\cite{kn:CC77} to
collect a set of structures at each program point $P$.
This involves finding the least fixed point of a certain set of
equations.
To ensure termination, the analysis is carried out with respect
to a finite abstract domain, that is,
the set of different structures is finite.
When the fixed point is reached, the structures that have
been collected at program point $p$ describe a superset of all the
concrete stores that can occur at $p$.
To determine whether a query is always satisfied at $p$, one checks
whether it holds in all of the structures that were collected there.
Instantiations of this framework are capable of establishing
nontrivial properties of programs that perform complex pointer-based
manipulations of {\em a priori\/} unbounded-size heap-allocated data
structures.
    %preliminaries
\section{Characterizing $3$-Valued Structures by First-Order Formulas}
\label{Se:Classic}

This section presents our results on characterizing $3$-valued structures
using first-order formulas.
Given a $3$-valued structure $S$, the question that we
wish to answer is whether it is possible to
give a formula $\gammaHat(S)$ that
accepts exactly
the set of $2$-valued structures that $S$ represents, i.e.,
$\C{S} \models \gammaHat(S)$ iff $\C{S} \in \gamma(S)$.

This question has different answers depending on what assumptions are made.
The task of generating a characteristic formula for a $3$-valued structure $S$ is challenging
because we have to find a formula
that identifies when embedding is possible,
i.e., that is satisfied by exactly those $2$-valued structures that embed into $S$.
It is not always possible to characterize
an \emph{arbitrary} $3$-valued structure by a first-order formula,
i.e.,
there exists a $3$-valued structure $S$ for which there is no
first-order formula with transitive closure that accepts exactly the set of
$2$-valued structures $\gamma$(S).

\begin{figure}
\begin{center}
\framebox{
\vbox{\xymatrix{
\xySummaryNode{u_1}\ar@{<.>}[r] & \xySummaryNode{u_2}\\
\xySummaryNode{u_3}\ar@{<.>}[u] \ar@{<.>}[ur] }}}
\end{center}
\caption{\label{Fi:Color}A $3$-valued structure that
represents $3$-colorable undirected graphs.
A $2$-valued structure can be embedded into this
structure if and only if it can be colored using $3$
colors.
}
\end{figure}
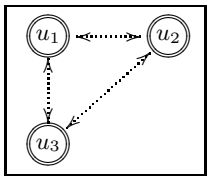

For example, consider the $3$-valued structure $S$ shown in \figref{Color}.
The absence of a self loop on any of the three summary nodes
implies that a $2$-valued structure can be embedded into this
structure if and only if it can be colored using $3$
colors (\lemref{embedding3color} in the appendix).
It is well-known that there exists no first-order formula,
even with transitive closure, that
expresses
$3$-colorability of undirected graphs,
unless $P = NP$ (e.g., see~\cite{Book:Immerman98,InCollection:Courcelle96}).\footnote{In fact,
the condition is even stronger. First-order
logic with transitive closure
can only express non-deterministic logspace (NL) computations,
thus, the NP-complete problem of $3$-colorability is not expressible in first-order logic,
unless $NL = NP$.
It is shown in~\cite{Book:Immerman98} using an ordering relation on the nodes.
In our context, without the ordering, the logic is less expressive.
Thus, the condition under which 3-colorability is expressible is even stronger than $NL = NP$.
We believe that there is an example of a $3$-valued structure that is not expressible in
the logic, independently of the question whether $P=NP$.
However, it is not the main focus of the current paper.}
Therefore, there is no first-order formula that accepts exactly the set $\gamma(S)$.
%???
%\footnote{Another example
%of a $3$-valued structure that cannot be characterized by a first-order formula with transitive closure
%can be derived by considering lists of even length and applying~cite{ChandraHarel82}.???}

\subsection{FO-Identifiable Structures}\label{Se:FOIdentStruct}

Intuitively, the difficulty in characterizing $3$-valued structures
is how to uniquely identify the
correspondence between concrete and abstract nodes using a first-order formula.
Fortunately, as we will see,
for the subclass of $3$-valued structures used in shape analysis
(also known as ``bounded structures''), the correspondence
can be easily defined using
first-order
%propositional
formulas.
The bounded structures are a subclass of the $3$-valued structures
in which it is possible to identify uniquely each node using a first-order formula.
%To avoid structures like the one shown in \figref{Color},
%We now formalize the notion of FO-identifiability,
%which generalizes condition for a
%subclass of $3$-valued structures, in which it is possible to
%uniquely identify each node using a formula.
\begin{definition}\label{De:UniqueFONames}
A $3$-valued structure $S$ is called {\bf FO-identifiable\/} if for every
node $u \in U^S$ there exists a first-order formula $\nodeFormula{S}{u}(w)$ with designated free variable $w$
such that
for every $2$-valued structure $\C{S}$ that embeds into $S$ using a function $f$,
for every concrete node $\C{u} \in U^{\C{S}}$ and for every node $u_i \in U^S$:
\begin{equation}
f(\C{u})= u_i
\iff
\C{S},[w \mapsto \C{u}] \models \nodeFormula{S}{u_i}(w)
\end{equation}
%$S$ is called FO-identifiable if all the nodes in $S$ are FO-identifiable.
%We say that a node $u$ in a $3$-valued structure $S$ is
%{\bf FO-identifiable\/} if there exists a first-order formula
%$\nodeFormula{S}{u}(w)$ with designated free variable $w$ such that for every
%$2$-valued structure $\C{S}$ that embeds into
%$S$ using a function $f$, and for every concrete node $\C{u} \in U^{\C{S}}$:
%\begin{equation}
%f(\C{u})= u
%\iff
%\C{S},[w \mapsto \C{u}] \models \nodeFormula{S}{u}(w)
%\end{equation}
%$S$ is called FO-identifiable if all the nodes in $S$ are FO-identifiable.
\end{definition}
The idea behind this definition is to have a formula that uniquely
identifies each node $u$ of the 3-valued structure $S$.
This will be used to identify the set of nodes of a $2$-valued
structure that are mapped to $u$ by embedding.
In other words, a concrete node $\C{u}$ satisfies the $node$ formula of
at most one abstract node, as formalized by the lemma:
\begin{lemma}\label{Lem:unique}
Let $S$ be an FO-identifiable structure, and let $u_1, u_2 \in S$ be distinct nodes.
Let $\C{S}$ be a 2-valued structure that embeds into $S$ and let $\C{u} \in \C{S}$.
At most one of the following hold:
\begin{enumerate}
\item $\C{S},[w \mapsto \C{u}] \models \nodeFormula{S}{u_1}(w)$
\item $\C{S},[w \mapsto \C{u}] \models \nodeFormula{S}{u_2}(w)$
\end{enumerate}
\end{lemma}
\begin{Remark}
\defref{UniqueFONames} can be generalized to handle arbitrary $2$-valued
structures, by also allowing extra designated free variables
for every concrete node and using equality to check
if the concrete node is equal to the designated variable:
$\nodeFormula{S}{u_i}(w, v_1, \ldots, v_n) \eqdef w = v_i$.
However, the equality formula cannot be used to identify
nodes in a $3$-valued structure because equality evaluates to $1/2$
on summary nodes.
\end{Remark}

We now introduce a standard concept for turning valuations into
formulas.
\begin{definition}\label{De:pb}
For a predicate $p$ of arity $k$ and truth value $B \in
\{0, 1, 1/2\}$, we define the formula $p^B(v_1, v_2, \ldots, v_k)$
to be the {\bf characteristic formula of $B$ for $p$\/},  by
\[
\begin{array}{lcl}
p^0(v_1, v_2, \ldots, v_k) & \eqdef & \neg p(v_1, v_2, \ldots, v_k)\\
p^1(v_1, v_2, \ldots, v_k) & \eqdef & p(v_1, v_2, \ldots, v_k) \\
p^{1/2}(v_1, v_2, \ldots, v_k) & \eqdef & 1\\
\end{array}
\]
\end{definition}

The main idea in the above definition is that,
for $B \in \{0,1\}$, $p^B$ holds when the value of $p$ is $B$, and
for $B = 1/2$ the value of $p$ is unrestricted.
\TrOnly{
This is formalized by the following lemma:
\begin{lemma}\label{Lem:Trivial}
For every $2$-valued structure $\C{S}$ and assignment $Z$
\[
\C{S}, Z \models p^B(v_1, \ldots, v_k) ~\mbox{iff}~
\iota^{\C{S}}(p)(Z(v_1), \ldots, Z(v_k)) \sqsubseteq B
\]
\end{lemma}
}

\defref{UniqueFONames} is not a constructive definition, because the
premises range over arbitrary $2$-valued structures and
arbitrary embedding functions.
For this reason, we now introduce a testable condition that implies FO-identifiability.

\para{Bounded Structures.}
The following subclass of $3$-values structures was defined in
\cite{kn:SRW99};\footnote{
This definition of bounded structures was given in
\cite{kn:SRW99}; it is slightly more restrictive than the one given
in~\cite{TOPLAS:SRW02,Thesis:LevAmi00}, which did not impose requirement \ref{De:BoundedStructures}(ii).
However, it does not limit the set of problems
handled by our method, if the structure that is bounded in the weak sense is also FO-identifiable.} the
motivation there was to guarantee that shape analysis was carried
out with respect to a finite set of abstract structures,
and hence that the analysis would always terminate.
\begin{definition}\label{De:BoundedStructures}
A {\bf bounded structure\/} over vocabulary $\Voc$ is a structure
$S = \B{U^S, \iota^S}$ such that for every $u_1, u_2 \in U^S$,
where $u_1 \neq u_2$, there exists a predicate symbol
$p \in \Voc_1$ such that (i)~$\iota^S(p)(u_1) \neq
\iota^S(p)(u_2)$ and (ii)~both $\iota^S(p)(u_1)$ and
$\iota^S(p)(u_2)$ are not $1/2$.
%In the sequel,
%$\BSTRUCT{\Voc}$ denotes the set of such structures.
\end{definition}
Intuitively, for each pair of nodes in a bounded structure,
there is at least one predicate that has different definite values for these nodes.
Thus, there is a finite number of different bounded structures (up to isomorphism).
%\begin{Remark}\label{lessBoundedRmark}
%This definition of bounded structures was given in
%\cite{kn:SRW99}; it is slightly more restrictive than the one given
%in~\cite{TOPLAS:SRW02,Thesis:LevAmi00}, which did not impose requirement \ref{De:BoundedStructures}(ii).
%However, it does not limit the set of problems
%handled by our method.
%Let $S$ be a $3$-valued structure that only
%satisfies the first requirement.
%It is possible to construct from
%$S$ a finite set of bounded structures $X$ such that
%$\gamma(X) = \gamma(S)$, using new unary predicates.
%This is based on the idea that
%a structure with an indefinite unary predicate value on a
%particular node $u$, can be represented by two (or
%three\footnote{when $u$ is a summary node we need to create an
%additional structure with two occurrences of $u$, for $0$ and $1$
%values for the predicate.}) structures with $0$ and $1$ values on
%$u$, respectively.

The following lemma shows that bounded structures are FO-identifiable
%using propositional formulas:
using formulas over unary predicates only (denoted by $\Voc_1$):
\begin{lemma}\label{Lem:Bounded}
Every bounded $3$-valued structure $S$ is FO-identifiable , where
\begin{equation}
\nodeFormula{S}{u_i}(w)  \eqdef \Land_{p \in \Voc_1} %AbstractionPredicates}
p^{\iota^{S}(p)(u_i)}(w)
\label{eq:individualFormula}
\end{equation}
\end{lemma}

\begin{example}\label{Ex:foRecognizable}
The first-order $node$ formulas for the structure $S$ shown in
\figref{threeStruct}, are:
\begin{eqnarray*}
\nodeFormula{S}{u_1}(w) = &
x(w) \land r_x(w) \land \neg y(w) \land \neg t(w) \land \neg e(w) \\
& \land \neg r_y(w) \land \neg r_t(w) \land \neg r_e(w) \land \neg is(w) \\
\nodeFormula{S}{u_2}(w) = &
\neg x(w) \land r_x(w) \land \neg y(w) \land \neg t(w) \land \neg e(w) \\
& \land \neg r_y(w) \land \neg r_t(w) \land \neg r_e(w) \land \neg is(w)
%\nodeFormula{S}{u_1}(w) & = &
%\neg is(w) \land x(w) \land \Land_{q \in \PVar-\{x\}} \neg q(w) \nonumber\\
%&& \land r_x(w) \land \Land_{q \in \PVar-\{x\}} \neg r_q(w) \\
%%%
%\nodeFormula{S}{u_2}(w) & = &
%\neg is(w) \land \Land_{q \in \PVar} \neg q(w) \nonumber\\
%&& \land r_x(w) \land \Land_{q \in \PVar-\{x\}} \neg r_q(w)
%%%
\end{eqnarray*}
\end{example}

\begin{Remark}
In the case that $S$ is a bounded $2$-valued structure,
the definition of a bounded structure becomes trivial.
The reason is that every node in $S$ can be named by
a quantifier-free formula built from unary predicates.
This is essentially the same as saying that every node can be named by a constant.
If structure $S'$ embeds into $S$, then $S'$ must be isomorphic to $S$,
therefore it is possible to name all nodes of $S'$ by the same constants.
However, this restricted case is not of particular interest for us,
because, to guarantee termination, shape analysis operates on structures
that contain summary nodes and indefinite values.
In the case that $S$ contains a summary node, a structure $S'$ that embeds into $S$ may
have an unbounded number of nodes;
hence the nodes of $S'$ cannot be named by a finite set of constants in the language.
\end{Remark}

We already know of interesting cases of FO-identifiable structures that are not bounded,
which can be used to generalize the abstraction defined in \cite{kn:SRW99}:

\begin{example}\label{Ex:FOIdentNotBounded}
The $3$-valued structure $S'$ in \figref{FOIdentNotBounded}
is FO-identifiable by:
\[
\begin{array}{ll}
\nodeFormula{S'}{u_1}(w) \eqdef &
x(w) \land r_x(w) \land \neg y(w) \land \neg t(w) \land \neg e(w) \\
& \land \neg r_y(w) \land \neg r_t(w) \land \neg r_e(w) \land \neg is(w) \\

\nodeFormula{S'}{u_2}(w) \eqdef &
\underline{\exists w_1 : x(w_1) \land n(w_1, w)} \land \neg x(w) \land r_x(w) \land \neg y(w) \land \neg t(w) \land \neg e(w) \\
& \land \neg r_y(w) \land \neg r_t(w) \land \neg r_e(w) \land \neg is(w) \\

\nodeFormula{S'}{u_3}(w) \eqdef &
\neg(\exists w_1 : x(w_1) \land n(w_1, w)) \land
\neg x(w) \land r_x(w) \land \neg y(w) \land \neg t(w) \land \neg e(w) \\
& \land \neg r_y(w) \land \neg r_t(w) \land \neg r_e(w) \land \neg is(w)
\end{array}
\]
However, $S'$ is not a bounded structure because nodes $u_2$ and $u_3$ have the same values of
unary predicates. To distinguish between these nodes, we extended $\nodeFormula{S'}{u_2}(w)$
with the underlined subformula, which captures the fact that only $u_2$ is directly pointed to by an $n$-edge
from $u_1$.
%\footnote{
%\begin{changebar}
%In general, every FO-identifiable structure can be transformed into a bounded structure by
%adding a new instrumentation predicate for each $node$ formula.
%\end{changebar}
%}
\begin{figure}
\begin{center}
%\framebox{
\begin{tabular}{|@{\hspace{1ex}}c@{\hspace{1ex}}|}
\hline
\xymatrix@R9pt@C12pt{
   %\xyEdge &
   &
     \xyNonSummaryNode{u_1}
             %\xyDottedLabeledEdge{n}
             \ar[r]^{n}
    &
    \xyNonSummaryNode{u_2}
             %\xyDottedLabeledEdge{n}
             \ar@{.>}[r]^{n}
    &
           \xySummaryNode{u_3} \ar@{.>}@(ur, ul)[]|{n} \\
   &             \tx, r_{x}\ar[u] &
                    r_{x}\ar[u] &
                    r_{x}\ar[u]
}\\
(S')\\
\hline
\end{tabular}
\end{center}
\caption{\label{Fi:FOIdentNotBounded}
A $3$-valued structure $S'$ is FO-identifiable, but not bounded.
}
\end{figure}
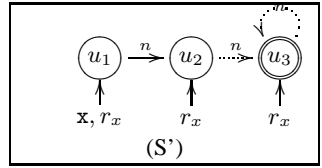
\end{example}
It can be shown that every FO-identifiable structure can be converted into a
bounded structure by introducing more instrumentation predicates.
For methodological reasons, we use the notion of FO-identifiable which directly capture
the ability to uniquely identify embedding via (FO) formulas.\footnote{In subsequent sections,
we redefine this notion to capture other classes of structures.}
One of the interesting features of FO-identifiable structures
is that the structures generated by a common TVLA operation ``focus'',
defined in \cite{Thesis:LevAmi00}, are all FO-identifiable (see \lemref{focusFOIdent} in \appref{Proofs}).
For example, \figref{FOIdentNotBounded}
shows the structure $S'$, which is
one of the structures resulting from applying the``focus'' operation
to the structure $S$ from \figref{threeStruct}(d)
with the formula $\exists v_1, v_2 : x(v_1) \land n(v_1, v_2)$.
$S'$ is FO-identifiable, but not bounded.
However, structures like the one shown in \figref{Color}
are not FO-identifiable unless $P = NP$.
%even if $P=NP$.
%- methodological: it simplifies the explanation and gives insight
%to when the abstraction can be characterized by FO+TC.
%- practical: the analysis is exponential in the number of predicates,
%thus, we prefer to work with "local" node predicates, defined for each structure,
%instead of defining the m as instrumentation predicates throughout the analysis.

\subsection{Characterizing FO-identifiable structures}\label{Se:FOFormula}

To characterize an FO-identifiable $3$-valued structure,
we must ensure
\begin{enumerate}
\item the existence of a surjective embedding function.
\item that every concrete node is represented by some abstract node.
\item that corresponding concrete and abstract predicate values meet the embedding condition
of \equref{EmbeddingCondition}.
\end{enumerate}

\begin{definition}\label{De:Characteristic}
\begin{Name}First-order Characteristic Formula\end{Name}
Let $S=\B{U=\{u_1, u_2, \ldots, u_n\}, \iota}$ be an FO-identifiable $3$-valued
structure.

We define the {\bf totality characteristic formula\/} to be the
closed formula:
\begin{equation}
\characteristicFormula{S}_{total} \eqdef \forall w:
        \Lor_{i=1}^n
        \nodeFormula{S}{u_i}(w)
\label{eq:ontoFormula}
\end{equation}

We define the {\bf nullary characteristic formula\/} to be the
closed formula:
\begin{equation}
\characteristicFormula{S}_{nullary} \eqdef \Land_{p \in \Voc_0}
p^{\iota^{S}(p)()} \label{eq:nullaryFormula}
\end{equation}

For a predicate $p$ of arity $r \geq 1$, we define the {\bf
predicate characteristic formula\/} to be the closed formula:
\begin{eqnarray}
\lefteqn{\characteristicFormula{S}[p] \eqdef \forall w_1, \ldots,
w_r: \Land_{\{u'_{1}, \ldots, u'_{r}\} \in U}} \nonumber\\
& \Land_{j=1}^r \nodeFormula{S}{u'_j}(w_j) \implies
p^{\iota^S(p)(u'_{1}, \ldots, u'_{r})}(w_1, \ldots, w_r)
\label{eq:naryFormula}
\end{eqnarray}

%Let $l$ be the largest arity of a predicate in $\Voc$.
The {\bf characteristic formula of $S$\/} is defined by:
\begin{equation}\label{eq:characteristicFormula}
\begin{array}{lll}
\characteristicFormula{S} & \eqdef &
    \Land_{i=1}^n (\exists v: \nodeFormula{S}{u_i}(v)) \\
    %(\exists v_1, \ldots, v_n: \Land_{i=1}^n \nodeFormula{S}{u_i}(v_i)
    %                    \land \Land_{k \neq j} \neg eq(v_k, v_j) ) \\
%    \left ( \begin{array}{l} \exists v_1, \ldots, v_n:\\
%                    \Land_{i=1}^n \nodeFormula{S}{u_i}(v_i)
%                        \land \Land_{k \neq j} \neg eq(v_k, v_j) \end{array} \right ) \\

     &\land  & \characteristicFormula{S}_{total}\\
     &\land  &    \characteristicFormula{S}_{nullary}\\
     &\land  &      \Land_{r=1}^{maxR} \Land_{p \in \Voc_r}
    \characteristicFormula{S}[p]
\end{array}
\end{equation}

The {\bf characteristic formula of set $X \subseteq  \STRUCT{\Voc}$\/}
is defined by:
\begin{equation}\label{eq:gammaHatSet}
\widehat{\gamma}(X) = F \land (\Lor_{S \in X} \characteristicFormula{S})
\end{equation}

Finally, for a singleton set $X = \{S\}$ we write $\widehat{\gamma}(S)$ instead of $\widehat{\gamma}(X)$.

\end{definition}

The main ideas behind the four conjuncts of \equref{characteristicFormula} are:
\begin{itemize}
  \item
    The existential quantification in the first conjunct
    requires that the $2$-valued structures
    have at least $n$ distinct nodes.
    For each abstract node in $S$, the first sub-formula locates
    the corresponding concrete node.
    Overall, this conjunct guarantees that
    embedding is surjective.
%    For each concrete individual, the first sub-formula locates
%    the corresponding individual in $S$.
%    Moreover, the first formula requires that these
%    individuals are represented by the corresponding individuals in
%    $S$ and are different.
  \item
    The totality formula ensures that every concrete node is
    represented by some abstract node.
    It guarantees that the embedding function is well-defined.
  \item
    The nullary characteristic formula ensures that the values of nullary
    predicates in the $2$-valued structures are at least as precise as
    the values of the corresponding nullary predicates in $S$.
  \item
    The predicate characteristic formulas guarantee that predicate
    values in the $2$-valued structures obey the requirements imposed by
    an embedding into $S$.\footnote{\defref{Characteristic} relates to all FO-identifiable structures,
    not only to bounded structures. For bounded structures, it can be simplified by omitting
    $\characteristicFormula{S}[p]$ for all unary predicates $p$, because it is implied
    by  $\characteristicFormula{S}_{total}$. In fact, it can be omitted only for the abstraction predicates, as defined
    in \cite{TOPLAS:SRW02}; however throughout this paper we consider all unary predicates to be abstraction predicates.}
\end{itemize}

\begin{example}\label{Ex:charFormula}
After a small amount of simplification, the characteristic formula $\widehat{\gamma}(S)$
for the structure $S$ shown in \figref{threeStruct}
is $F_{List} \land \characteristicFormula{S}$, where
$\characteristicFormula{S}$ is:
\[
\begin{array}{@{\hspace{0ex}}r@{\hspace{1.0ex}}l@{\hspace{0ex}}}
          & \exists v: \nodeFormula{S}{u_1}(v) \land \exists v: \nodeFormula{S}{u_2}(v)\\
    \land & \forall w: \nodeFormula{S}{u_1}(w) \lor \nodeFormula{S}{u_2}(w)\\
    \land & \Land_{p \in \Voc_1} \forall w_1:\Land_{i=1,2}
           (\nodeFormula{S}{u_i}(w_1)
           \implies p^{\iota^S(p)(u_i)}(w_1))\\
    \land & \begin{array}[t]{@{\hspace{0ex}}r@{\hspace{1.0ex}}l@{\hspace{0ex}}}
              \forall w_1, w_2: & (\nodeFormula{S}{u_1}(w_1) \land \nodeFormula{S}{u_1}(w_2)\implies %\\
                                eq(w_1, w_2) \land \neg n(w_1, w_2) \land \neg n(w_2, w_1)) \\
                          \land & (\nodeFormula{S}{u_1}(w_1) \land \nodeFormula{S}{u_2}(w_2)\implies %\\
                                \neg eq(w_1, w_2) \land \neg n(w_2, w_1))
            \end{array}
\end{array}
\]
The $node$ formulas are given in \exref{foRecognizable}, and the
predicates for the {\tt insert} program in \figref{Insert}(b) are
shown in \tableref{Predicates}.
Above, we simplified the formula from \equref{characteristicFormula} by combining
implications that had the same premises.
The integrity formula $F_{List}$ is given in \exref{IntegrityExample}.
Note that it uses transitive closure to define the reachability predicates;
consequently, $\widehat{\gamma}(S)$ is a formula in first-order logic with transitive closure.
\end{example}

When a predicate has an indefinite value on some node tuple,
a corresponding conjunct of \equref{naryFormula} can be omitted, because it simplifies to $\TRUE$.

Thus, the size of this simplified version of $\characteristicFormula{S}$ is
linear in the number of definite values of predicates in $S$.
Assuming that the $node^S$ formulas contain no quantifiers or transitive-closure operator,
e.g., when $S$ is bounded,
the $\characteristicFormula{S}$ formula has no quantifier alternation, and does not contain
any occurrences of the transitive-closure operator.
Thus, the formula $\widehat{\gamma}$ is in Existential-Universal normal form
(and thus decidable for satisfiability) whenever $F$ is in Existential-Universal normal form
and does not contain transitive closure.\footnote{
For practical reasons, we often replace the $node$ formula by a new (definable) predicate,
and add its definition to the integrity formula.
%Similarly, we can show that \defref{UniqueFONames}
%does not allow transitive-closure operator in $node$ formula,
}
Moreover, if the maximal arity of the predicate in $\Voc$ is $2$,
then $\gammaHat$ is in the two-variable fragment of first-order logic~\cite{Mortimer75}, wherever
$F$ is.
In \secref{Applications}, we discuss other conditions under which $\gammaHat$ can be expressed in a
decidable logic.

The following theorem shows that for every FO-identifiable structure $S$,
the formula $\widehat{\gamma}(S)$
accepts exactly the set of $2$-valued structures represented by $S$.
\begin{theorem}\label{The:RepresentingStructuresByFormulae}
For every FO-identifiable $3$-valued structure $S$,
and $2$-valued structure $\C{S}$,
$\C{S} \in \gamma(S)~ \mbox{iff}~\C{S} \models
\widehat{\gamma}(S)$.
\end{theorem}
      %from 3-v strcut to FO formula
\section{Supervaluational Semantics for First-Order Formulas}\label{Se:Supervaluation}

In this section, we consider the problem of how to extract
information from a $3$-valued structure by evaluating a query.
A compositional semantics for $3$-valued first-order logic is defined
in \cite{TOPLAS:SRW02};
however, that semantics is not as precise as the one defined here.
The semantics given in this section can be seen as providing the limit
of obtainable precision.

\para{The Notion of Supervaluational Semantics} %\label{Se:SpecificationSupervaluation}
defined below,
has been used in \cite{JPhil:vanFraassen66,CONCUR:BG00}.
%generalizes \cite{JPhil:vanFraassen66}.
%,LICS:RLS02}.

\begin{definition}\label{De:Supervaluation}
\begin{Name}Supervaluational Semantics of First-Order Formulas\end{Name}
Let $X$ be a set of $3$-valued structures and $\varphi$ be a closed
formula.
The {\bf supervaluational semantics of $\varphi$ in $X$\/},
denoted by $\superval{\varphi}(X)$, is defined to be the join of
the values of $\varphi$ obtained from each of the $2$-valued structures
that $X$ represents, i.e., the most-precise conservative
value that can be reported for the value of formula $\varphi$ in
the $2$-valued structures represented by $X$ is
\begin{equation}
  \label{Eq:SpecificationOfSupervaluational1}
  \superval{\varphi}(X) =
     \left\{\begin{array}{l@{\hspace{.4cm}}l}
       1    & \mbox{{\rm if\/}~} \C{S} \models \varphi~\mbox{{\rm for all\/}~} \C{S} \in \gamma(X) \\
       0   & \mbox{{\rm if\/}~} \C{S} \not\models \varphi~\mbox{{\rm for all\/}~} \C{S} \in \gamma(X) \\
       1/2 & \mbox{{\rm otherwise\/}~}
     \end{array}\right.
\end{equation}
\end{definition}

%Also notice that the embedding theorem
%implies that $\superval{\varphi}(S) \sqsubseteq
%\threeValue{\varphi}{S}{[]}$. It is possible to define
%supervaluational semantics for formulas with free variables.

The compositional semantics given in~\cite{TOPLAS:SRW02} and used in TVLA
can yield $1/2$ for $\varphi$, even when the value of $\varphi$
%${\semp{\varphi}}_2^{\C{S}}$
is $1$ for all the $2$-valued structures $\C{S}$ that $S$ represents
(or when the value of $\varphi$
is $0$ for all the $\C{S}$).
In contrast, when the supervaluational semantics yields $1/2$,
we \emph{know} that any sound extraction of information from
$S$ must return $1/2$.

\begin{example}\label{Ex:SupExample}
We demonstrate now that the supervaluational semantics of
the formula
$\varphi_{ {\tt x\rightarrow next \neq NULL} } \eqdef \exists v_1, v_2: x(v_1) \land n(v_1, v_2)$
on the structure $S$
from \figref{threeStruct}(d) is $1$.
That is, we wish to argue that for all of the $2$-valued structures that
structure $S$ from \figref{threeStruct}(d) represents,
the value of the formula
$\varphi_{ {\tt x\rightarrow next \neq NULL} }$
must be $1$.

We reason as follows: $S$ represents
a list with at least two nodes;
i.e., all $2$-valued structures represented by $S$ have at least two nodes.
One node, $u^\natural_1$, corresponding to $u_1$ in $S$,
is pointed to by program variable {\tt x}.
The other node, corresponding to the summary node $u_2$, must be
reachable from \tx. Consider the sequence of nodes reachable from \tx,
starting with $u^\natural_1$.
Denote the first node in the sequence that embeds into $u_2$
by $u^\natural_2$. By the definition of reachability,
there must be an \tn-link to $u^\natural_2$ from a node embedded into $u_1$.
But the integrity rules guarantee that there is exactly one node
that embeds into $u_1$, namely, $u^\natural_1$.
Therefore, the formula $x(v_1) \land n(v_1, v_2)$
holds for $[v_1 \mapsto u^\natural_1, v_2 \mapsto u^\natural_2]$.

Note that this formula will be evaluated to $1/2$ by TVLA,
because $x(v_1) \land n(v_1, v_2)$ evaluates to $1/2$ under the assignment
$[v_1 \mapsto u_1, v_2 \mapsto u_2]$:
the compositional semantics yields
$x(u_1) \land n(u_1, u_2) = 1 \land 1/2 = 1/2$.
\end{example}

Notice that \defref{Supervaluation} does not provide a constructive
way to compute $\superval{\varphi}(X)$ because $\gamma(X)$ is
usually an infinite set.

\para{Computing Supervaluational Semantics using Theorem Provers.} %\label{Se:ImplementationSupervaluation}
If an appropriate theorem prover is at hand,
$\superval{\varphi}(S)$ can be computed with
the procedure shown in \figref{ImplementationOfSupervaluational}.
This procedure is not an algorithm, because the theorem prover might not terminate.
Termination can be assured by using standard techniques (e.g., having the theorem prover return a
safe answer if a time-out threshold is exceeded) at the cost of losing
the ability to guarantee that a most-precise result is obtained.
If the queries posed by operation {\tt Supervaluation} can be expressed in
a decidable logic, the algorithm for computing supervaluation can be implemented
using a decision procedure for that logic.
In \secref{Applications}, we discuss such decidable logics that are useful for shape analysis.
\begin{figure}
\begin{center}
\framebox{
\begin{minipage}{1in}
\begin{alltt}
\begin{tabbing}
proc\=edure Supervaluation(\=$\varphi$: Formula,\+\+\\
    X: Set of $3$-valued structures):  Value\-\\
if ($\gammaHat(X) \implies \varphi$ is valid) return $1$;\\
else if ($\gammaHat(X) \implies \neg \varphi$ is valid) return $0$;\\
otherwise return $1/2$;
\end{tabbing}
\end{alltt}
\end{minipage}
}
\end{center}
\caption{\label{Fi:ImplementationOfSupervaluational}A procedure for computing
the supervaluational value of a formula $\varphi$ that
encodes a query on a $3$-valued structures $S$.
}
\end{figure}
%\begin{equation}
%  \label{eq:ImplementationOfSupervaluational}
%  \superval{\varphi}(S) =
%     \left\{\begin{array}{l@{\hspace{.4cm}}l}
%       1    & \mbox{{\rm if\/}~} \gammaHat(S) \implies \varphi~\mbox{{\rm is valid}} \\
%       0   & \mbox{{\rm if\/}~} \gammaHat(S) \implies \neg\varphi~\mbox{{\rm is valid}} \\
%       1/2 & \mbox{{\rm otherwise\/}~}
%     \end{array}\right.
%\end{equation}
     %supervaluation
\section{Applications}\label{Se:Applications}

% Executive view
The experiments discussed in this section demonstrate how
the $\gammaHat$ operation can be harnessed in the context of program analysis:
the results described below go beyond what previous systems were capable of.
In \secref{SPASS}, we discuss the use existing theorem provers and their limitations.
In \secref{EADTC}, we suggest a way to overcome these limitations, using decidable logic.

We present two examples that use $\gammaHat$ to read out
information from $3$-valued structures in a conservative, but rather precise way.
The first example demonstrates how supervaluational semantics allows us to obtain more precise
information from a $3$-valued structure than we would have using compositional semantics. %otherwise.
The second example demonstrates how to use the $3$-valued structures obtained from a TVLA analysis
to construct a loop invariant; this is then used to show
that certain properties of a linked data structure hold on each
loop iteration.
In addition, we briefly describe how $\gammaHat$ can be used in algorithms for computing most-precise
abstraction operations for shape analysis.
Finally, we report on other work that employs $\gammaHat$ to generate a
concrete counter-example for shape analysis.

\begin{Remark}
The $\gammaHat$ operation defines a symbolic concretization with respect to a given abstract domain.
In~\secref{Classic}, we defined $\gammaHat$ for the abstract domain of sets of $3$-valued
structures. %ordered by the Hoare ordering on sets, where structures are ordered by the embedding ordering.
In \appref{Canonic}, we describe a related abstract domain and define $\gammaHat$ for it.
The applications described in this section can be used with any domain
for which $\gammaHat$ is defined in some logic and a theorem prover for that logic exists.
In our examples, we use $\gammaHat$ defined in \secref{Classic} and
the first-order logic with transitive closure.
\end{Remark}

%\subsection{Implementation Issues}\label{Se:SPASS}
\subsection{Using the First-Order Theorem Prover SPASS}\label{Se:SPASS}

%% What is TVLA and what was done
The TVLA (\cite{SAS:LS00}) system performs an iterative fixed-point computation, which
yields at every program point $p$ a set $X_p$ of bounded structures.
It guarantees that $\gamma(X_p)$ is a superset of the $2$-valued structures that can
arise at $p$ in any execution.
%To carry out these experiments,
We have implemented the $\gammaHat$ operation in TVLA, and
employed SPASS~\cite{SPASS} to check, using the formula $\gammaHat(X_p)$,
that certain properties of the heap hold at program point $p$.
Also, we implemented the supervaluational procedure described in \secref{Supervaluation},
employing SPASS.
The enhanced version of TVLA generates the formula
$\gammaHat(S)$ and makes at most two calls to SPASS
to compute the supervaluational value of a query $\varphi$
in structure $S$.
In this section, we report on our experience in using SPASS and the problems we have encountered.

%SPASS
First, calls to SPASS theorem prover need not terminate, because
first-order logic is undecidable in general.
However, in the examples described below, SPASS always terminated.

\begin{example}
In \exref{SupExample} we (manually) proved that the supervaluational value of
the formula $\varphi_{ {\tt x\rightarrow next \neq NULL} }$ on the structure $S$
from \figref{threeStruct}(d) is $1$.
To check this automatically, we used SPASS to determine the validity of
$\widehat{\gamma}(S) \implies \varphi_{ {\tt x\rightarrow next \neq NULL} }$;
SPASS indicated that the formula is valid.
This guarantees that the formula $\varphi_{ {\tt x\rightarrow next \neq NULL} }$
evaluates to $1$ on all of the $2$-valued structures that embed into $S$.

In contrast, TVLA uses Kleene semantics for $3$-valued formulas, and
will evaluate $\varphi_{ {\tt x\rightarrow next \neq NULL} }$ to $1/2$:
under the assignment $[v_1 \mapsto u_1, v_2 \mapsto u_2]$,
$x(v_1) \land n(v_1, v_2)$ evaluates to $1 \land 1/2$, which equals $1/2$.
\end{example}

\subsubsection{Generating and Querying a Loop Invariant}\label{Se:loopExperiment}
We used TVLA to compute, for each program point $p$,
a set $X_p$ of bounded structures that overapproximate
the set of stores that may occur at that point.
We then generated $\widehat{\gamma}(X_p)$.
Because TVLA is sound,
$\widehat{\gamma}(X_p)$ must be an invariant
that holds at program point $p$, according to \theref{RepresentingStructuresByFormulae}.
In particular, when $p$ is a program point that begins a loop,
$\widehat{\gamma}(X_p)$ is a loop invariant.

\begin{example}\label{Ex:loopInvariantEx}
Let $X = \{S_i \mid i=1,\ldots,5\}$ denote
the set of five $3$-valued structures that TVLA
found at the beginning of the loop in the
\tinsert\/ program from \figref{threeStruct}.
\tableref{StructFormula} and \tableref{StructFormulaTwo}
of \appref{AppendixTable} show the $S_i$ and their
characteristic formulas.
The loop invariant is defined by
\[
  \widehat{\gamma}(X) = F_{List} \land (\Lor_{i=1}^5 \characteristicFormula{S_i})
\]

Using SPASS, this formula was then used to check that in every
structure that can occur at the beginning of the loop, \tx\/ points
to a valid list, i.e., one that is acyclic and unshared.
This property is defined by the following formulas:
\[
\begin{array}{@{\hspace{0ex}}l@{\hspace{.5ex}}l@{\hspace{.5ex}}l@{\hspace{0ex}}}
  \text{acyc}_x & \eqdef & \forall v_1, v_2: r_x(v_1) \land n^+(v_1, v_2) \implies  \neg n^+(v_2, v_1) \\
  \text{uns}_x  & \eqdef & \forall v \colon r_x(v) \implies
                ~\neg (\exists w_1, w_2 \colon \neg eq(w_1, w_2) \land n(w_1, v) \land n(w_2, v)) \\
  \text{list}_x & \eqdef & \text{acyc}_x \land \text{uns}_x
\end{array}
\]
We applied SPASS to check the validity of $\gammaHat(S) \implies list_x$; SPASS indicated that
the formula is valid.\footnote{SPASS input is available from {\tt www.cs.tau.ac.il/$\sim$gretay}.}
\end{example}

In addition to the termination issue, a second obstacle is that SPASS considers infinite structures, which
are not allowed in our setting.\footnote{Our
intended structures are finite, because they represent memory configurations,
which are guaranteed to be finite, although their size is not bounded.}
As a consequence, SPASS can fail to verify that a formula is valid for
our intended set of structures;
however, the opposite can never happen: whenever SPASS indicates that
a formula is valid, it is indeed valid for our intended set of
structures.
\begin{example}
We tried to verify that every concrete linked-list represented by the $3$-valued structure
$S$ from \figref{threeStruct}(d)
has a last element.
This condition is expressed by the formula
$\varphi_{last} \eqdef \exists v_1 \forall v_2 : \neg n(v_1, v_2)$.
The supervaluational value of $\varphi_{last}$ on a structure $S$ is
$\superval{\varphi}(S) = 1$, for the following reasons.
Because $r_x$ has the definite value $1$ on $u_2$ in $S$,
all concrete nodes represented by the summary node $u_2$
must be reachable from $x$.
Thus, these nodes must form a linked list, i.e.,
each of these concrete nodes, except for one node that is the ``last'',
has an $n$-edge to another concrete node represented by $u_2$.
The last node does not have an $n$-edge back to any of the nodes represented by $u_2$,
because that would create sharing, whereas the value of predicate $is$ in $S$ is $0$ on $u_2$.
Also, the last node cannot have an $n$-edge to the concrete node represented by $u_1$,
because the value of predicate $n$ on the pair $\B{u_2, u_1}$ in $S$ is $0$.
Therefore, the last element cannot have an outgoing $n$-edge.

We used SPASS to determine the validity of
$\gammaHat(S) \implies \varphi_{last}$;
SPASS indicated that the formula is \emph{not} valid, because it considered
a structure that has infinitely many concrete nodes, all represented by $u_2$.
Each of these concrete nodes has an $n$-edge to the next node.

The validity test of the formula $\gammaHat(S) \implies \neg \varphi_{last}$
failed, of course, because there exists a finite structure that is represented by $S$
(and thus satisfies $\gammaHat(S)$) and has a last element. For example, the structure
in \figref{threeStruct}(a) that represents a list of size $2$.
Therefore, the procedure \textit{Supervaluation}$(\varphi_{last}, {S})$ implemented using SPASS
returns $1/2$, even though the supervaluational value of $\varphi_{last}$ on $S$ is $1$.
\end{example}

The third, and most severe, problem that we face is that SPASS does not support transitive closure.
Because transitive closure is not expressible in first-order logic,
we could only partially model transitive closure in SPASS,
%integrity rules,
as described below.

SPASS follows other theorem provers in allowing axioms to express
requirements on the set of structures considered.
We used SPASS axioms to model integrity rules.
To partially model transitive closure, %integrity rules,
we replaced uses of $n^+(v_1, v_2)$ by uses of a new designated
predicate $t[n](v_1, v_2)$.
Therefore, SPASS will consider some structures that do not represent
possible stores.
As a consequence, SPASS can fail to verify that a formula is valid for
our intended set of structures;
however, the opposite can never happen: whenever SPASS indicates that
a formula is valid, it is indeed valid for our intended set of
structures.
To avoid some of the spurious failures to prove validity, we added
axioms to guarantee that
(i)\ $t[n](v_1, v_2)$ is transitive and
(ii)\ $t[n](v_1, v_2)$ includes all of $n(v_1, v_2)$;
thus, $t[n](v_1, v_2)$ includes all of $n^+(v_1, v_2)$.
Because transitive closure requires a minimal set, which is not
expressible in first-order logic, this approach provides a
looser set of integrity rules than we would like.
However, it is still the case that whenever SPASS indicates that a
formula is valid, it is indeed valid for the set of structures in
which $t[n](v_1, v_2)$ is exactly $n^+(v_1, v_2)$.

\begin{example}
SPASS takes into account the structure shown in \figref{SPASS_TC},
in which the value of $t[n](u_1, u_3)$ is $1$,
but the value of $n^+(u_1, u_3)$ is $0$
because there is no $n$-edge from $u_2$ to $u_3$.
\begin{figure}
\begin{center}
%\framebox{
\begin{tabular}{|@{\hspace{1ex}}c@{\hspace{1ex}}|}
\hline
\xymatrix@R9pt@C8pt{
     \xyNonSummaryNode{u_1}
             \ar[rr]^{n}
             \ar@/^0.8pc/[rr]^{t[n]}
             \ar@/^2.5pc/[rrr]^{t[n]}
    &&
           \xyNonSummaryNode{u_2}
      &
           \xyNonSummaryNode{u_3}  \\
                   \tx, r_{x}\ar[u] &&
                    r_{x}\ar[u] &
                    r_{x}\ar[u]
}\\
\hline
\end{tabular}
\end{center}
\caption{\label{Fi:SPASS_TC} SPASS takes into account structures in which the $t[n]$ predicate
overapproximates the  $n^+$ predicate, such as the structure shown in this figure.}
\end{figure}
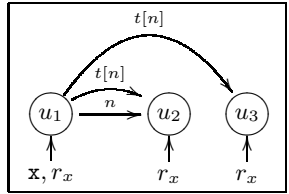
\end{example}

\subsection{Decidable Logic}\label{Se:EADTC}
%% Decidable logics

The obstacles mentioned in \secref{SPASS} are not specific to SPASS.
They occur in all theorem provers for first-order logic that we are aware of.
To address these obstacles, we are investigating the use of a decidable logic.
To reason about linked data structures, we need a notion of reachability
to be expressible, for example, using transitive closure.
However, a logic that is both decidable and includes reachability
must be limited in other aspects.
%Because this notion tends to make logics undecidable, e.g., \cite{GOR99,lics:IRRSY},
%only limited logics with reachability are decidable.

One such example is the decidable second-order theory of two successors \textit{WS2S} \cite{Rabin};
its decision procedure is implemented in a tool called MONA \cite{KlaEtAl:Mona}.
Second-order quantification suffices to express reachability,
but there are still two problems.
First, the decision procedure for \textit{WS2S} is necessarily non-elementary \cite{Meyer}.
Second, \textit{WS2S} only applies to trees, or,
equivalently, to function graphs (graphs with at most one edge leaving any vertex).

Another example is $\Lone$, which is a subset of first-order logic with transitive closure,
in which the following restriction are imposed on formulas:
(i)~they must be in existential-universal form,
and (ii)~they must use at most a single unary function $f$, but can use an arbitrary number of unary predicates.
\cite{csl04:IRRSY} shows that the decision procedure for satisfiability of $\Lone$ is NEXPTIME-complete.

In spite of their limitations, both \textit{WS2S} and $\Lone$ can be useful for
reasoning about shape invariants and mutation operations on data structures, such as
singly and doubly linked lists, (shared) trees, and graph types \cite{kn:KS93}.
The key is the \textit{simulation technique} \cite{cav04:IRRSY}, which encodes
complex data-structures using \emph{tractable} structures,
e.g., function graphs or simple trees, where we can reason with decidable logics.
%The idea is to detect that a mutation is to violate
%the correctness of the simulation and forbid such mutations
%from being applied.

For example, given a suitable simulation,
$\gammaHat$ formula can be expressed in \textit{WS2S} and $\Lone$ if the integrity formula $F$ can.
This follows from the definition of $\gammaHat$ in \equref{gammaHatSet} and the fact
that $\characteristicFormula{S}$ does not contain
quantifier alternation.
This makes $\Lone$ and \textit{WS2S} candidate implementations
for the decision procedure used in the supervaluational semantics and in the algorithms
described below.

%\cite{lics:IRRSY} describes the logic $\Lone$ that is a subset of first-order logic with transitive closure,
%imposed by the following limitations on a $\Lone$ formula: (i)~existential-universal form,
%and (ii)~a single unary function $f$ and an arbitrary number of unary predicates.
%
%In spite of its limitations, in \cite{cav:IRRSY} the logic $\Lone$ is shown to be useful for
%reasoning about shape invariants and mutation operations on data structures, such as
%singly and doubly linked lists, (shared) trees, and graph types \cite{kn:KS93}.
%The key is the \textit{simulation technique} that encodes
%complex data-structures using $\Lone$ logic, including mutations of these data-structures.
%It detects that a mutation is to violate the correctness of the simulation and forbids such mutations
%from being applied.
%
%Also, the satisfiability of $\Lone$ formulas is decidable and NEXPTIME-complete,
%hence the $\Lone$ decision procedure is a candidate implementation for the {\it isSatifiable}
%function.\footnote{
%Another candidate is the decision procedure for monadic 2-nd order logic over trees \cite{KlaEtAl:Mona}, MONA,
%which has non-elementary complexity.}

%\subsection{Querying Using Supervaluational Semantics}\label{Se:supExperiment}

\subsection{Assume-Guarantee Shape Analysis} % using $\Lone$ Logic}
\label{Se:Assume}
%%Assume algorithm
The $\gammaHat$ operation is useful beyond computing supervaluational semantics: it is a
necessary operation used in the algorithms described in~\cite{TACAS:YRS04,VMCAI:RSY04}.
These algorithms perform abstract operations symbolically by representing abstract values
as logical formulas, and use a theorem prover to check validity of these formulas.
These algorithms improve on existing shape-analysis techniques by:
\begin{itemize}
\item conducting abstract interpretation in the most-precise fashion,
improving the technique used in the TVLA system~\cite{SAS:LS00,TOPLAS:SRW02},
which provides no guarantees about the precision of its basic mechanisms.
\item performing modular verification using assume-guarantee reasoning and procedure specifications.
This is perhaps the most-exciting potential application of $\gammaHat$ (and $\Lone$ logic),
because existing mechanisms for shape analysis, including TVLA, do not support
assume-guarantee reasoning.
\end{itemize}
%For example,
%given a suitable simulation of vocabularies \cite{cav:IRRSY},
%$\gammaHat$ can be expressed in $\Lone$ if the integrity formula $F$ can.
%It follows from the definition of $\gammaHat$ in \equref{gammaHatSet} and the fact
%that $\characteristicFormula{S}$ is in $\Lone$, because it does not contain
%quantifier alternation and transitive closure.

%These algorithm is based on multiple queries.
%The simulation technique \cite{cav:IRRSY} allows us to translate these queries to a decidable logic,
%and utilize a decision procedure for that logic to implement the algorithms.
%The main part of the translation involves translating $\gammaHat$ formulas to a decidable logic.
%For example, $\gammaHat(S)$ can be expressed in $\Lone$ if the integrity rules $IR$ can.
%The reason is that $\gammaHat(a) \eqdef \Land_{S \in a} \xi^S \land IR$, and
%$\xi$ is in $\Lone$ because it does not contain
%quantifier alternation and transitive closure.
%In \cite{lics:IRRSY} we provide a sufficient condition under which this can be done automatically.
%

\subsection{Counter-example Generation}\label{Se:CounterExample}

%The techniques presented in this paper have already been explored in order to improve the applicability of
%TVLA~\cite{Unpublished:ESY03,Master:Erez04}.
Some preliminary work to use the techniques presented in this paper to improve the applicability of TVLA
has been carried out.
The tool described in ~\cite{Unpublished:ESY03,Master:Erez04} uses the $\gammaHat$ operation to generate a concrete counter-example for
a potential error message produced by TVLA
for an intermediate $3$-valued structure $S$ at a program point $p$.
Such a tool is useful to check if a reported error is a real error or a false-alarm,
i.e., it never occurs on any concrete store.

Generation of concrete counter-examples from $S$ proceeds as follows.
First, $S$ is converted to the formula $\gammaHat(S)$.
Then, the tool uses weakest precondition to generate a formula that represents the stores at the entry
point that lead to an execution trace that reaches $p$ with a store that satisfies $\gammaHat(S)$.
%Then, the tool enumerates the traces starting at the program entry that lead
%to a concrete store in $S$ by computing the weakest
%precondition which guarantees that the trace is executed with a resultant store which
%satisfies the generated formula holds.
Finally, a separate tool \cite{MACE} generates a concrete store that satisfies the
formula for the entry point.
\section{Related Work}
\label{Se:RelatedWork}

There is a sizeable literature on {\em structure-description formalisms\/}
for describing properties of linked data structures (see
\cite{esop:BRS99,TOPLAS:SRW02} for references).
The motivation for the present paper was to understand the expressive
power of the shape abstractions defined in \cite{TOPLAS:SRW02}.

In previous work, Benedikt et al.\ \cite{esop:BRS99} showed how to
translate two kinds of shape descriptors, ``path matrices''
\cite{kn:Hendren,kn:HN90} and the variant of shape graphs
discussed in \cite{kn:SRW98}, into a logic called $L_r$
(``logic of reachability expressions'').
The shape graphs from \cite{kn:SRW98} are also amenable to the
techniques presented in the present paper:
the characteristic formula defined in \equref{characteristicFormula}
is much simpler than the translation to $L_r$ given in \cite{esop:BRS99};
moreover, \equref{characteristicFormula} applies to a more
general class of shape descriptors.
However, the logic used in \cite{esop:BRS99} is decidable, which
guarantees that terminating procedures can be given for
problems that can be addressed using $L_r$.

The Pointer Analysis Logic Engine (PALE) \cite{PLDI:MS01}
provides a structure-description formalism that serves
as an assertion language; assertions are translated to
second-order monadic logic and fed to MONA.
PALE does not handle all data structures, but can handle
all data structures describable as graph types \cite{kn:KS93}.
Because the logic used by MONA is decidable, PALE
is guaranteed to terminate.

One point of contrast between the shape abstractions
based on $3$-valued structures studied in this paper
and both $L_r$ and the PALE assertion language is
that the powerset of $3$-valued structures forms
an abstract domain.
This means that $3$-valued structures can be used for program analysis
by setting up an appropriate set of equations and finding its fixed
point \cite{TOPLAS:SRW02}.
In contrast, when PALE is used for program analysis,
an invariant must be supplied for each loop.

\TrOnly{Other structure-description formalisms in the literature
include ADDS \cite{kn:HHN92} and shape types \cite{kn:FL97}.}

\TrOnly{The supervaluational semantics for first-order logic discussed in
\secref{Supervaluation} is related to a number of other
supervaluational semantics for partial logics and $3$-valued logics
discussed in the literature
\cite{JPhil:vanFraassen66,HPL:Blamey2Ed,CONCUR:BG00}.
%,LICS:RLS02}.
Compared to previous work, an innovation of
\figref{ImplementationOfSupervaluational} is the use of $\gammaHat$
to translate a $3$-valued structure to a formula.
In fact, \figref{ImplementationOfSupervaluational} is an example
of a general reductionist strategy for providing a
supervaluational evaluation procedure for abstract domains by using existing
logics and theorem-provers/decision-procedures.
%;as we discuss at greater length in \cite{VMCAI:RSY04},
%a supervaluational evaluation procedure can be obtained
%whenever an appropriate logic, $\gammaHat$ function, and
%theorem-prover/decision-procedure are available.
}
 % related work

A recent work~\cite{VMCAI04:KR}, which is an abbreviated version of a more extensive presentation of the results
reported in \cite{closureTR:KM03}, provides
an alternative characterization of $3$-valued structures using logical formulas,
equivalent to the characterization presented in the present paper.
The present paper, which extends and elaborates on the results of
\cite{Thesis:Yorsh03}, unlike \cite{VMCAI04:KR,closureTR:KM03},
reports on experience and algorithmic issues in using
logical characterization of structures for shape analysis;
this material is important because
shape analysis is the primary motivation and the intended application of this paper,
as well as \cite{VMCAI04:KR,closureTR:KM03}.
Also, \secref{closure} of the present paper gives a simple semantic argument for
the property of closure under negation, shown in \cite{closureTR:KM03} using a different formalism.
The technical similarities and differences between the two works are described
in a note available from {\tt www.cs.tau.ac.il/$\sim$gretay}.

\section{Final Remarks}
\label{Se:FinalRemarks}

In \cite{VMCAI:RSY04}, we discuss how
to perform all operations required for abstract interpretation in the
most-precise way possible (relative to the abstraction in use),
if certain primitive operations can be carried out,
and if a sufficiently powerful theorem prover is at hand.
Chief among the primitive operations that must be available is $\gammaHat$;
thus, the material that has been presented in this paper shows
how to fulfill the requirements of \cite{VMCAI:RSY04}
for a family of abstractions based on $3$-valued structures
(essentially those used in our past work \cite{TOPLAS:SRW02}
and in the TVLA system \cite{SAS:LS00}).

In ongoing work, we are investigating the feasibility of actually
applying the techniques from \cite{VMCAI:RSY04} to perform
abstract interpretation for abstractions based on $3$-valued
structures.
This approach could be more precise than TVLA because, for instance,
it would take into account in a first-class way the integrity formula
of the abstraction.
In contrast, in TVLA some operations temporarily ignore the integrity
formula, and rely on later clean-up steps to rectify matters.

Another step can be taken in this direction,
which is to eliminate the use of $3$-valued structures,
and directly carry out fixed-point computations over logical formulas.

We are also investigating the feasibility of using the results from
this paper to develop a more precise and modular version of TVLA by
using \emph{assume-guarantee} reasoning \cite{TACAS:YRS04}.
The idea is to allow arbitrary first-order formulas with transitive
closure to be used to express pre- and post-conditions, and to
analyze the code for each procedure separately.

\subsubsection*{Acknowledgements} We thank Neil Immerman, Viktor Kuncak, Tal Lev-Ami, and Alexander Rabinovich
for their contributions to this paper.

%\begin{itemize}
%  \item
%    The above process becomes more attractive in the presence of
%    complex integrity formulas, which are ignored in the compositional
%    algorithm. This provides an alternative to coerce and to focus.
%  \item
%    To use TVLA as a generator for loop invariants for existing
%    verification systems.
%  \item
%    To provide a ``search space'' for systems like \cite{POPL:FQ02}.
%  \item
%    To check the invariants produced by TVLA.
%\end{itemize}

%\end{spacing}

{\small
%\begin{spacing}{0.9}
  \bibliography{df,logic,more,mab,static}
  \bibliographystyle{plain}
%\end{spacing}
}

\appendix
\section{Characterizing Canonical Abstraction by First-Order Formulas}\label{Se:Canonic}

This section defines an alternative abstract domain for use in shape analysis
(and other logic-based analyses).
This domain keeps more explicit information than the one in \secref{Embedding}
and enjoys nice closure properties (see~\secref{closure}).
This domain uses a particular class of embedding functions that are defined by
%This section provides an alternative formulation of the abstraction which
%does not rely on embedding, as is the case for the abstraction defined in \secref{Embedding}.
%Instead, the new abstract domain relies on
a simple operation,
called {\em canonical abstraction}, which maps $2$-valued structures into
a limited subset of bounded structures.

\subsection{Canonical Abstraction}
%We define {\em tight embedding\/},
%in which information loss is minimized when multiple nodes of
%$S$ are mapped to the same node in $S'$:
%
%\begin{definition}\label{De:Abstraction}
%A structure $S' = \B{U^{S'}, \iota^{S'}}$ is a {\bf tight
%embedding\/} of $S = \B{U^S, \iota^S}$ if
%there exists a surjective function $\blur \colon U^S \to U^{S'}$
%such that, for every $p \in \Voc_k$ of arity $k$,
%\begin{equation}
%  \iota^{S'}(p)(u'_1,  \ldots, u'_k) = \\
%  \bigsqcup_{
%    {\scriptsize
%    \begin{array}{@{\hspace{0ex}}c@{\hspace{0ex}}}
%      u_i \in U^{S}, \mbox{{\rm s.t.}} \\
%      \blur(u_i) = u'_i \in U^{S'}, \\
%      {1 \le i \le k}
%    \end{array}}
%  }
%  \iota^S(p)(u_1, \ldots, u_k)
%  \label{eq:abstraction}
%\end{equation}
%
%When a surjective function $\blur$ possesses
%property from ~\equref{abstraction}, we say that $S' = \blur(S)$.
%\end{definition}

Canonical abstraction was defined in \cite{kn:SRW99} as an abstraction with
the following properties:
\begin{itemize}
\item It provides a uniform way to obtain $3$-valued structures
of a priori bounded size.
This is important to automatically derive properties
of programs with loops by employing iterative
fixed-point algorithms.
Canonical abstraction maps concrete nodes into
abstract nodes according to
the definite values of the unary predicates.
\item The information loss is minimized when multiple nodes of
$S$ are mapped to the same node in $S'$,
\end{itemize}
This is formalized by the following definition:
\begin{definition}\label{De:CanonicalAbstraction}
A structure $S' = \B{U^{S'}, \iota^{S'}}$ is
a {\bf canonical abstraction\/} of a structure $S$,
if $S \sqsubseteq^{\blur} S'$, where $\blur \colon U^S \to U^{S'}$
is the following surjective mapping:
\begin{equation}\label{eq:tightEmbed}
  \blur(u) = u_{\{ p \in \Voc_1 \mid \iota^{S}(p)(u) = 1 \},
               \{ p \in \Voc_1  \mid \iota^{S}(p)(u) = 0 \}
              }
\end{equation}
and, for every $p \in \Voc_k$ of arity $k$,
\begin{equation}\label{eq:canonicName}
  \iota^{S'}(p)(u'_1,  \ldots, u'_k) = \\
  \bigsqcup_{
    {\scriptsize
    \begin{array}{@{\hspace{0ex}}c@{\hspace{0ex}}}
      u_i \in U^{S}, \mbox{{\rm s.t.}} \\
      \blur(u_i) = u'_i \in U^{S'}, \\
      {1 \le i \le k}
    \end{array}}
  }
  \iota^S(p)(u_1, \ldots, u_k)
\end{equation}
We say that $S' = \blur(S)$.
\end{definition}

The name ``$u_{\{ p \in \Voc_1 \mid \iota^{S}(p)(u) = 1 \},
               \{ p \in \Voc_1 \mid \iota^{S}(p)(u) = 0 \}}$''
is known as the {\bf canonical name\/} of node $u$.
The subscript on the canonical name of $u$ involves two sets of
unary predicate symbols:
(i)~those that are true at $u$, and
(ii)~those that are false at $u$.

\begin{example}\label{Ex:CanonicalAbsInsertNaive}
  In structure $S$ from \figref{threeStruct},
  the canonical names of the nodes are as follows:
  \begin{center}
    \scalebox{.9}{
    \begin{tabular}{|c|l|}
      \hline
        {\bf Node\/} & {\bf Canonical Name\/} \\
      \hline
        $u_1$ & $u_{\{x, r_x\},\{y, t, e, is, r_y, r_t, r_e\}}$ \\
        $u_2$ & $u_{\{r_x\},\{x, y, t, e, is, r_y, r_t, r_e\}}$ \\
      \hline
    \end{tabular}
    }
  \end{center}
In the context of canonical abstraction, $S$ shown in \figref{threeStruct}
represents $S_b$ and $S_c$, but not $S_a$; i.e.,
$S$ represents lists that are pointed to by $x$ that have
at least \textit{three} nodes, but it does not represent a list with just two nodes.
The reason is that predicates $n$ and $eq$ have indefinite values in $S$,
but a list with only two nodes cannot have both $0$ and $1$ values for
the corresponding entries, as required for minimizing information loss as
defined in \equref{canonicName}.\footnote{\equref{canonicName} is called the \textit{tight-embedding} condition in \cite{TOPLAS:SRW02}.}
In contrast, according to the abstraction that relies on embedding, defined in \secref{Embedding},
$S$ represents lists with \textit{two} or more elements.
\end{example}

To characterize canonical abstraction, we define the set of $3$-valued structures that
are ``images of canonical abstraction'' (\ICA),
i.e., the results of applying canonical abstraction to $2$-valued structures.
%To characterize canonical abstraction, we have to redefine
%the subset of $3$-valued structures of interest.
%We are interested only in $3$-valued structures that
%that are ``images of canonical abstraction'' (\ICA),
%i.e., results of applying canonical abstraction to $2$-valued structures.
\begin{definition}\label{De:ICA}
\begin{Name}Image of canonical abstraction (\ICA)\end{Name}
Structure $S$ is an \ICA\ if there exists a $2$-valued structure $\C{S}$ such that
$S$ is the canonical abstraction of $\C{S}$.
\end{definition}

\para{Concretization of $3$-Valued Structures.}
Canonical abstraction allows us to define the (potentially infinite)
set of $2$-valued structures represented by a set of $3$-valued structures, that are \ICA\:
\begin{definition}\label{De:ConcreteCanonic}
\begin{Name}Concretization of ICA %$3$-Valued
Structures\end{Name}
For a set of structures $X \subseteq  \STRUCT{\Voc}$, that are \ICA\ structures,
we denote by $\gamma_c(X)$ the set of $2$-valued structures that $X$ represents,
i.e.,
\begin{equation} \label{eq:CanonicConcretization}
\begin{array}{r}
  \gamma_c(X) = \left \{ \begin{array}{l}
                          \C{S} \in \TSTRUCT{\Voc} \mid
                             \text{exists}~ S \in X \text{such that}\\
                             S~\text{is the canonical abstraction of }\C{S}~\text{and}~\C{S} \models F
                    \end{array}  \right \}
\end{array}
\end{equation}
Also, for a singleton set $X = \{S\}$ we write $\gamma_c(S)$ instead of $\gamma_c(X)$.
\end{definition}

The abstract domain is the powerset of \ICA\ structures, where the order relation is set inclusion.
Note that this abstract domain is finite, because there is a finite  number of different ICA structures (up to isomorphism).
Denote by $\alpha_c$ the extension of the abstraction function $\blur$ to sets.
This defines a Galois connection $\B{\alpha_c, \gamma_c}$ between sets of $2$-valued structures and sets of \ICA\ structures.

\subsection{Canonical-FO-Identifiable Structures}

We define the notion of canonical-FO-identifiable nodes
using canonical abstraction rather than embedding, which was
used for the notion of FO-identifiable nodes in \defref{UniqueFONames}.
\begin{definition}\label{De:CanonicUniqueFONames}
We say that a node $u$ in a $3$-valued structure $S$
is {\bf canonical-FO-identifiable\/} if there exists a formula
$\nodeFormula{S}{u}(w)$ with designated free variable $w$,
such that for every $2$-valued structure $\C{S}$,
if $S$ is the canonical abstraction of $\C{S}$, i.e., $\C{S} \in \gamma_c(S)$,
then for every concrete node $\C{u} \in U^{\C{S}}$:
\begin{equation}
\blur(\C{u})= u
\iff
\C{S},[w \mapsto \C{u}] \models \nodeFormula{S}{u}(w)
\end{equation}
$S$ is called canonical-FO-identifiable if all the nodes in $S$ are canonical-FO-identifiable.
\end{definition}

We can also prove \lemref{unique} for the case of canonical abstraction rather than embedding.

\subsection{Characterizing Canonical Abstraction}

An \ICA\ structure is always a bounded structure, in which all nullary and unary predicates
have definite values.\footnote{If not all unary predicates
are defined as abstraction predicates, then the result may be
a bounded structure of the less restrictive kind mentioned in \secref{FOIdentStruct}.
Also, unary predicates that are not abstraction predicates may have indefinite values.}
This is formalized by the following lemma:
\begin{lemma}\label{Lem:ICAProp}
If\ $3$-valued structure $S = \B{U^S, \iota^S}$ over vocabulary $\Voc$
is \ICA\, then:
\begin{description}
    \item [(i)] $S$ is a bounded structure.
    \item [(ii)] For each nullary predicate $p$, $\iota^S(p)() \in \{0,1\}$.
    \item [(iii)] For each element $u \in U$ and each unary predicate $p$, $\iota^S(p)(u) \in \{0,1\}$.
\end{description}
\end{lemma}

The following lemma shows that \ICA\ structures are canonical-FO-identifiable:
%have FO-identifiable nodes.
\begin{lemma}\label{Lem:CanonicBounded}
Every $3$-valued structure $S$ that is an \ICA\, is canonical-FO-identifiable, where
\begin{equation}
\nodeFormula{S}{u_i}(w)  \eqdef \Land_{p \in \Voc_1} %AbstractionPredicates}
p^{\iota^{S}(p)(u_i)}(w)
\label{eq:canonIndividualFormula}
\end{equation}
\end{lemma}

Using this fact, we can define a formula $\tau^S$ that
accepts exactly
the set of
$2$-valued structures represented by $S$ under canonical abstraction.
The formula $\tau^S$ is merely $\characteristicFormula{S}$
with additional conjuncts to ensure that the information loss is minimized,
i.e., for every predicate $p$ and every $1/2$ entry of $p$, the $2$-valued structure has both
a corresponding 1 entry and a corresponding 0 entry.

\begin{definition}\label{De:TightChar}
\begin{Name}First-Order Characteristic Formula for Canonical Abstraction\end{Name}
Let $3$-valued structure $S=\B{U^S, \iota}$ be an ICA.

For a predicate $p$ of arity $r$, we define the closed formula for $p$:
\begin{equation} \label{eq:tightnaryFormula}
%\begin{array}{rl}
%\begin{eqnarray}
\tau^{S}[p] \eqdef
\Land_{\scriptsize \begin{array}{c}
\{u'_{1}, \ldots, u'_{r}\} \subseteq U^S\\
\mbox{s.t. }\iota^S(p)(u'_{1}, \ldots, u'_{r}) = 1/2
\end{array}}
%\Land_{\{u'_{1}, \ldots, u'_{r}\} \in U^S}
%\Land_{\iota^S(p)(u'_{1}, \ldots, u'_{r}) = 1/2}\\
\left ( \begin{array}{rl}
& \exists w_1, \ldots, w_r : \Land_{j=1}^r \nodeFormula{S}{u'_j}(w_j) \land p(w_1, \ldots, w_r) \\
\land & \exists w_1, \ldots, w_r : \Land_{j=1}^r \nodeFormula{S}{u'_j}(w_j) \land \neg p(w_1, \ldots, w_r)
\end{array}
\right )
%\end{eqnarray}
%\end{array}
\end{equation}

The formula of $S$ is defined by:
\begin{equation}
\tau^S \eqdef \characteristicFormula{S} \land
    %\tau^S_{nullary} \land \tau^{S}_{unary} \land
    \Land_{r=2}^{maxR} \Land_{p \in \Voc_r}
    \tau^S[p]
\label{eq:tightChareteristicFormula}
\end{equation}

The {\bf characteristic formula for canonical abstraction
of a set of \ICA\/ structures} $X \subseteq \STRUCT{\Voc}$ is defined by
\begin{equation}\label{eq:gamma_c_X}
\widehat{\gamma}_c(X) = F \land (\Lor_{S \in X} \tau^{S})
\end{equation}
Also, for a singleton set $X = \{S\}$, where $S$ is an \ICA\ structure,
we write $\widehat{\gamma}_c(S)$ instead of $\widehat{\gamma}_c(X)$.
\end{definition}

\begin{example}\label{Ex:tightcharFormula}
The characteristic formula for canonical abstraction of the structure $S$ shown in
\figref{threeStruct}(d) is:
\begin{equation}\label{eq:tightcharex}
\begin{array}{rl}
\widehat{\gamma}_c(S) & = \widehat{\gamma}(S)\\
\land &  \exists w_1, w_2: \nodeFormula{S}{u_1}(w_1) \land \nodeFormula{S}{u_2}(w_2) \land n(w_1, w_2) \\
\land & \exists w_1, w_2: \nodeFormula{S}{u_1}(w_1) \land \nodeFormula{S}{u_2}(w_2) \land \neg n(w_1, w_2) \\
\land & \exists w_1, w_2: \nodeFormula{S}{u_2}(w_1) \land \nodeFormula{S}{u_2}(w_2) \land n(w_1, w_2) \\
\land & \exists w_1, w_2: \nodeFormula{S}{u_2}(w_1) \land \nodeFormula{S}{u_2}(w_2) \land \neg n(w_1, w_2) \\
\land & \exists w_1, w_2: \nodeFormula{S}{u_2}(w_1) \land \nodeFormula{S}{u_2}(w_2) \land eq(w_1, w_2) \\
\land & \exists w_1, w_2: \nodeFormula{S}{u_2}(w_1) \land \nodeFormula{S}{u_2}(w_2) \land \neg eq(w_1, w_2) \\
\end{array}
\end{equation}
where  $\widehat{\gamma}(S)$ is given in \exref{charFormula}.
As explained in \exref{CanonicalAbsInsertNaive}, $S$ does not represent
a list of two nodes; the corresponding $2$-valued structure $S_a$, shown in \figref{threeStruct}(a),
does not satisfy \equref{tightcharex},
because the last four lines cannot be satisfied by any assignment in $S_a$.
\end{example}
\begin{Remark}
The formula $\tau^S$ does not contain quantifier alternation and transitive closure.
Therefore, $\widehat{\gamma}_c$ is in Existential-Universal normal form (and thus decidable) whenever $F$
is in Existential-Universal form and does not contain transitive closure.
\end{Remark}

\begin{theorem}\label{The:RepresentingTightEmbeddingByFormulae}
For every $3$-valued structure $S$ that is an \ICA\,
and $2$-valued structure $\C{S}$
\[
%\mbox{S is a canonical abstraction of}~\C{S}
\C{S} \in \gamma_c(S)
~\mbox{iff}~\C{S} \models \widehat{\gamma}_c(S)
\]
\end{theorem}

\subsection{Closure Properties of ICA Structures}\label{Se:closure}

This section gives a simple semantic proof that
the class of formulas that characterize ICA structures is closed under negation.
This result was shown in \cite{closureTR:KM03} using a different formalism.

From \equref{canonicName} it follows that for two distinct ICA structures $S_1$ and $S_2$,
$\gamma_c(S_1) \cap \gamma_c(S_2) = \varnothing$.
Intuitively, each $2$-valued structure can be represented by exactly one ICA structure.
This implies that the complement of the concretization of an ICA structure
can be represented precisely by a finite set of ICA structures.

\newcommand{\domain}{\mathcal{D}}
\newcommand{\comp}{\mathcal{C}}
Denote by $\domain$ the set of all $2$-valued structures
that satisfy the integrity formula $F$:
$\domain \eqdef \{ \C{S} \in \TSTRUCT{\Voc} \mid \C{S} \models F \}$.
\begin{lemma}\label{Lem:noIntersection}
Let $S$ be an ICA structure.
There exists a set of ICA structures $X$ such that
$\gamma_c(X) = \domain \smallsetminus \gamma_c(S)$.
\end{lemma}
This can be reformulated using \theref{RepresentingTightEmbeddingByFormulae}
in terms of characteristic formulas for ICA structures.
This shows that the class of formulas that characterize canonical abstraction is
closed under negation, in the following sense:
\begin{lemma}\label{Lem:closedUnderNegation}
Consider the formula $\tau^S$ from \equref{tightChareteristicFormula}, for some ICA structure $S$.
There exists a set of ICA structures $X$, such that
the formula $F \land \neg \tau^S$ is equivalent to the formula $\gammaHat_c(X)$.
%Let $\varphi$ be a formula such that $\varphi \eqdef \gammaHat_c(S)$
%for some ICA structure $S$.
%There exists a set of ICA structures $X$, such that
%the formula $\neg \varphi$ is equivalent to the formula $\gammaHat_c(X)$.
\end{lemma}
\begin{Remark}
Note that \lemref{noIntersection} and \lemref{closedUnderNegation} do not hold for
bounded structures, described in \secref{FOIdentStruct}.
The reason, intuitively, is that some $2$-valued structures can be represented by more than one bounded structure.

For example, consider the $2$-valued structure $S_a$ from \figref{threeStruct}, which denotes a linked-list of length exactly $2$.
It is in the concretization of two different $3$-valued structures: the first is $S_a$ itself, considered as
a $3$-valued structure  $S'$ (that represents a single $2$-valued structure: $\gamma(S') = \{ S_a \}$);
the second is the structure $S$ from \figref{threeStruct}.

For the purpose of this example, assume that the integrity formula $F$ (that
defines $\domain$) requires that all elements be reachable from $x$,
in addition to the integrity formula $F_{List}$ from \exref{IntegrityExample}.
The complement $\comp \eqdef \domain \smallsetminus \gamma(S') = \domain \smallsetminus S_a$
is the set that contains an empty linked list, a linked list of length $1$, and
linked lists of length $3$ or more.
The representation of $\comp$ is a set $X$ of bounded structures.
To capture linked lists of length $3$ or more, $X$ must contain a $3$-valued structure $S$ from \figref{threeStruct}.
However, $\gamma(S)$ includes a list of length $2$ as well, denoted by $S_a$, which is not in $\comp$. Therefore, $\gamma(S) \neq \comp$.
\end{Remark}

\section{Characterizing General $3$-Valued Structures by NP Formulas}\label{Se:NPFormula}

In this section, we show how to characterize general $3$-valued structures.

\subsection{Motivating Example}

If the input structure is FO-identifiable, \theref{RepresentingStructuresByFormulae}
ensures that the result of operation $\gammaHat$ precisely captures the concretization of the input structure.
The purpose of this example is to show what happens if we apply the $\gammaHat$ operation,
as defined in \secref{Classic}, to a structure that is not FO-Identifiable.
When $S$ is not FO-identifiable, $\widehat{\gamma}(S)$
only provides a sufficient test for the embedding
of $2$-valued structures into $S$.
\begin{example}
The $3$-valued structure $S$ shown in \figref{Color}
describes undirected graphs. We draw undirected edges as
two-way directed edges.
This structure uses a set of predicates $\Voc = \{eq, f, b\}$,
where $f(v_1, v_2)$ and $b(v_2, v_1)$ denote
the forward and backward directions of an edge between
nodes $v_1$ and $v_2$.

When \equref{characteristicFormula} is applied to the $3$-valued structure $S$
shown in \figref{Color}, we get
%Let predicates $f$ and $b$ describe the undirected edges of
%the graph represented by $S$.
\begin{equation}\label{eq:FullNPExample}
\begin{array}{ll}
\Land_{i=1}^3 & \exists v : \nodeFormula{S}{u_i}(v) \\
                     \land & \forall w: \Lor_{i=1}^3 \nodeFormula{S}{u_i}(w) \\
                    \land & \forall w_1, w_2:
                    \Land_{k \neq j} (\nodeFormula{S}{u_k}(w_1) \land \nodeFormula{S}{u_j}(w_2) \implies
                    f^{1/2}(w_1, w_2))\\
                     \land & \forall w_1, w_2:
                    \Land_{k \neq j} (\nodeFormula{S}{u_k}(w_1) \land \nodeFormula{S}{u_j}(w_2) \implies
                    b^{1/2}(w_1, w_2))\\
                     \land & \forall w_1, w_2:
                    \Land_{i=1}^3( \nodeFormula{S}{u_i}(w_1) \land \nodeFormula{S}{u_i}(w_2) \implies b^0(w_1, w_2))\\
                     \land & \forall w_1, w_2:
                    \Land_{i=1}^3( \nodeFormula{S}{u_i}(w_1) \land \nodeFormula{S}{u_i}(w_2) \implies f^0(w_1, w_2))
\end{array}
\end{equation}
Because this example does not include unary predicates,
the $node$ formula given in \lemref{Bounded} evaluates to $\TRUE$
on all elements. Hence, \equref{FullNPExample} can be simplified to:
%\begin{equation}
\[
\begin{array}{lll}
&\Land_{i=1}^3 & \exists v : \TRUE \\
                    &\land & \forall w: \Lor_{i=1}^3 \TRUE \\
                    &\land & \forall w_1, w_2:
                    \Land_{k \neq j} (\TRUE \land \TRUE \implies \TRUE)\\
                    &\land & \forall w_1, w_2:
                    \Land_{k \neq j} (\TRUE \land \TRUE \implies \TRUE)\\
                    &\land & \forall w_1, w_2:
                    \Land_{i=1}^3( \TRUE \land \TRUE \implies \neg b(w_1, w_2))\\
                    &\land & \forall w_1, w_2:
                    \Land_{i=1}^3( \TRUE \land \TRUE \implies \neg f(w_1, w_2))
\end{array}
\]
%\end{equation}
After further simplification, we get the formula
$\forall w_1, w_2: \neg f(w_1, w_2)\land \forall w_1, w_2: \neg b(w_1, w_2)$.
The simplification is due to the fact that the implication in \equref{naryFormula}
unconditionally holds for all pairs of distinct nodes,
because $f$ and $b$ evaluate to $1/2$ on those pairs,
except for the requirement imposed by the absence of self-loops in $S$.

This formula is only fulfilled by graphs with %at least $3$ nodes and
no edges, which are obviously $3$-colorable. But this formula is
too restrictive: it does not capture some $3$-colorable graphs.

%\begin{changebar}
%It is well-known that $3$-colorability cannot be expressed
%with a first-order formula with transitive closure, unless $P = NP$,
%because it is an NP-complete problem~\cite{Book:Immerman98}.\footnote{First-order
%\begin{changebar}
%logic with transitive closure
%can only express non-deterministic logspace (NL) computations.
%It is shown in~\cite{Book:Immerman98} using an ordering relation on the nodes.
%In our context, without the ordering, the logic is less expressive.
%Thus, the condition under which 3-colorability is expressible is even stronger than $NL = NP$.
%We believe that an example of $3$-valued structure that is not expressible in
%the logic, independently of the question whether $P=NP$, can be given.
%However, it is not the main focus of the current paper.
%\end{changebar}}
%\end{changebar}
%It is not surprising that $3$-colorability cannot be expressed
%with a first-order formula with transitive closure,
%because it is an NP-complete problem
%and even with transitive closure, first-order logic can only
%express non-deterministic logspace computations.~\cite{Book:Immerman98}~\footnote{Note
%\begin{changebar}
%that~\cite{Book:Immerman98}
%uses an ordering on the nodes, and even with this ordering, first-order logic with transitive closure
%can only express non-deterministic logspace computations.
%\end{changebar}}
\end{example}

\subsection{Characterizing General $3$-Valued Structures}

Existential monadic second-order formulas are a subset of
Fagin's second-order formulas \cite{kn:Fagin75}, named NP formulas,
which capture NP computations.
A formula in existential monadic second-order logic has the form:
\[
  \exists V_1, V_2, \ldots, V_n: \varphi
\]
where the $V_i$ are set variables, and $\varphi$ is a first-order
formula that can use membership tests in $V_i$.
We show that in this subset of second-order logic,
the characteristic formula from \defref{Characteristic}
can be generalized to handle arbitrary $3$-valued
structures using existential quantification over set variables
(with one set variable for each abstract node).\footnote{
This result is mostly theoretical. In principle,
this encoding falls into monadic-second order logic,
which is decidable if we restrict the concrete structures of interest to trees.
However, we have not investigated this direction further.
}

\begin{definition}\label{De:NPCharacteristic}
\begin{Name}NP Characteristic Formula\end{Name}
Let $S=\B{U=\{u_1, u_2, \ldots, u_n\}, \iota}$ be a $3$-valued
structure.

We define the following formula to ensure that the sets are non\_empty:
\begin{equation}\label{eq:nonemptyFormula}
\NPcharacteristicFormula{S}_{non\_empty}[i] \eqdef \exists w_i: \nodeFormula{S}{u_i}(w_i)
\end{equation}

We define the following formula to ensure that the sets $V_k$, $V_j$
are disjoint:
\begin{equation}\label{eq:disjoinFormula}
\NPcharacteristicFormula{S}_{disjoint}[k, j] \eqdef
\forall w_1, w_2: \nodeFormula{S}{u_k}(w_1) \land \nodeFormula{S}{u_j}(w_2) \implies \neg eq(w_1, w_2)
\end{equation}

The {\bf NP characteristic formula of $S$\/} is defined by:
\begin{equation}\label{eq:NPcharacteristicFormula}
\begin{array}{@{\hspace{0ex}}l@{\hspace{1.0ex}}r@{\hspace{1.0ex}}l@{\hspace{0ex}}}
\NPcharacteristicFormula{S} & \eqdef
   \exists V_1, \ldots, V_n: & \Land_{i=1}^n \NPcharacteristicFormula{S}_{non\_empty}[i] \land %non-empty sets
     \Land_{k \neq j} \NPcharacteristicFormula{S}_{disjoint}[k, j]\\ %disjoint sets
     &\land  & \characteristicFormula{S}_{total}\\
     &\land  & \characteristicFormula{S}_{nullary}\\
     &\land  & \Land_{r=1}^{maxR} \Land_{p \in \Voc_r}
    \characteristicFormula{S}[p]
\end{array}
\end{equation}
where $\characteristicFormula{S}_{total}$, $\characteristicFormula{S}_{nullary}$,
$\characteristicFormula{S}[p]$ are defined as in \defref{Characteristic},
except that $\nodeFormula{S}{u_i}$ is the NP formula
$\nodeFormula{S}{u_i}(w) \eqdef (w \in V_i)$.
(Here, we abuse notation slightly by referring to $V_i$ in $\nodeFormula{S}{u_i}(w)$.
This could have been formalized by passing $V_1, \ldots, V_n$ as extra parameters
to $\nodeFormula{S}{u_i}$.)

The {\bf NP characteristic formula of a finite set $X \subseteq  \STRUCT{\Voc}$\/}
is defined by:
\begin{equation}
\gammaHat_{NP}(X) = F \land (\Lor_{S \in X} \NPcharacteristicFormula{S})
\end{equation}

Finally, for a singleton set $X = \{S\}$ we write $\gammaHat_{NP}(S)$ instead of $\gammaHat_{NP}(X)$.

\end{definition}

%\begin{example}
%After a small amount of simplification,
%the NP characteristic formula $\NPcharacteristicFormula{S}$ for the graph
%shown in \figref{Color} is:
%\[
%\begin{array}{@{\hspace{0ex}}l@{\hspace{1.0ex}}l@{\hspace{1.0ex}}l@{\hspace{0ex}}}
%  \exists V_1, V_2, V_3: && \Land_{i=1}^3 \exists w_i: \nodeFormula{S}{u_i}(w_i) \\
%    & \land & \Land_{k \neq j} \forall w_1, w_2: (\nodeFormula{S}{u_k}(w_1) \land \nodeFormula{S}{u_j}(w_2) \implies \neg eq(w_1, w_2)) \\ %ctive sets
%    & \land & \forall w: \Lor_{i=1}^3 \nodeFormula{S}{u_i}(w) \\ %total
%    & \land & \forall w_1,  w_2: \Land_{i=1}^3 (\Land_{j=1,2} \nodeFormula{S}{u_i}(w_j) \implies
%                                   \neg f(w_1,w_2) \land \neg b(w_1, w_2))\\
%    & \land & \forall w_1, w_2: \Land_{k \neq j} (\nodeFormula{S}{u_k}(w_1) \land \nodeFormula{S}{u_j}(w_2) \implies \neg eq(w_1, w_2))
%\end{array}
%\]
%and the $node$ formulas are given in \defref{NPCharacteristic}.
%\end{example}

\begin{example}
After a small amount of simplification,
the NP characteristic formula $\NPcharacteristicFormula{S}$ for the graph
shown in \figref{Color} is:
\[
\begin{array}{@{\hspace{0ex}}r@{\hspace{1.0ex}}l@{\hspace{1.0ex}}l@{\hspace{2.0ex}}c@{\hspace{0ex}}}
  \exists V_1, V_2, V_3: & \Land_{i=1}^3 (\exists w: w \in V_i) & (i) \\
     \land & \Land_{k \neq j} \forall w_1, w_2: ( w_1 \in V_k \land w_2 \in V_j \implies \neg eq(w_1, w_2)) & (ii) \\ %disjunctive sets
     \land & \forall w: \Lor_{i=1}^3 w \in V_i & (iii) \\ %total
     \land & \forall w_1,  w_2: \Land_{i=1}^3 (\Land_{j=1,2} w_j \in V_i \implies
                                   \neg e(w_1,w_2) \land \neg e(w_2, w_1)) & (iv)
% \\
%    & \land & \forall w_1, w_2: \Land_{k \neq j} (w_1 \in V_k \land w_2 \in V_j \implies \neg eq(w_1, w_2)) & (v)
\end{array}
\]
In this formula, $V_1$, $V_2$, and $V_3$ represent the three color classes.
Line by line, the formula says:
(i) each color class has at least one member;
(ii) the color classes are pairwise disjoint;
(iii) every node is in a color class;
(iv) nodes in the same color class are not connected by an undirected edge.
\end{example}

The following theorem generalizes the result in \theref{RepresentingStructuresByFormulae}
for an arbitrary $3$-valued structure $S$, using
NP-formula $\gammaHat_{NP}(S)$ to
accept exactly
the set of $2$-valued structures represented by $S$.
\begin{theorem}\label{The:RepresentingStructuresByNPFormulae}
For every $3$-valued structure $S$,
and $2$-valued structure $\C{S}$:
\[
\C{S} \in \gamma(S)~ \mbox{iff}~\C{S} \models
\gammaHat_{NP}(S)
\]
\end{theorem}
      %from general 3-v structure to SO formula
\section{Generating and Querying a Loop Invariant}\label{Se:AppendixTable}

\tableref{StructFormula} and \tableref{StructFormulaTwo} show the structures and the
characteristic formulas for the experiment described in \exref{loopInvariantEx}.

It is interesting to note that the size of $\xi^{S_2}$ is bigger than the size of $\xi^{S_1}$.
This is natural because $S_2$ has more definite values, which impose more restrictions than
are imposed by $S_1$.
\begin{table}
\centering
\resizebox{\textwidth}{!}{
$\begin{array}{|l|l|}
\hline
{\bf Structure\/} & {\bf Characteristic Formula\/}\\
\hline
  \xymatrix@R10pt@C10pt{
   \tx,\ty\ar[r]          &  
        \xyNonSummaryNode{u_1}\ar@{.>}[r]^n &
        \xySummaryNode{u_2}\ar@{.>}@(ur, ul)[]|{n}\\
S_1	&r_x, r_y \ar[u] & r_x, r_y \ar[u]
}

 &
\begin{array}{lll}
%% S_1
%%% nodes
%%
\nodeFormula{S_1}{u_1}(w)&=& x(w) \land y(w) \land \neg t(w) \land \neg e(w)\\
& \land & r_x(w) \land r_y(w) \land \neg r_t(w) \land \neg r_e(w) \land \neg is(w)\\
\nodeFormula{S_1}{u_2}(w) & =& \neg x(w) \land \neg y(w) \land \neg t(w) \land \neg e(w) \\
& \land & r_x(w) \land r_y(w) \land \neg r_t(w) \land \neg r_e(w) \land \neg is(w)\\
%% formula
\hline
\characteristicFormula{S_1}& = & \Land_{i=1,2} (\exists v: \nodeFormula{S_1}{u_i}(v))\\
    & \land & \forall w: \Lor_{i=1,2} \nodeFormula{S_1}{u_i}(w)\\
    & \land & \forall w_1,  w_2: \Land_{i=1,2} \nodeFormula{S_1}{u_i}(w_i) \implies \\
    && \neg eq(w_1, w_2) \land \neg n(w_2, w_1) \\
    & \land & \forall w_1,  w_2: \Land_{i=1,2} \nodeFormula{S_1}{u_1}(w_i) \implies \\
    && \land eq(w_1, w_2) \land \neg n(w_1, w_2) \\
\end{array}
\\
\hline
  \xymatrix@R10pt@C10pt{
   \tx\ar[r]          &  
        \xyNonSummaryNode{u_1}\ar[r]^n &
        \xyNonSummaryNode{u_2}\\
S_2	&r_x \ar[u] & y, r_x, r_y \ar[u]
}

   &
  \begin{array}{lll}
%% S_2
%%% nodes
%%
\nodeFormula{S_2}{u_1}(w)&=& x(w) \land \neg y(w) \land \neg t(w) \land \neg e(w) \\
& \land & r_x(w) \land \neg r_y(w) \land \neg r_t(w) \land \neg r_e(w) \land \neg is(w) \\
\nodeFormula{S_2}{u_2}(w) & =& \neg x(w) \land \neg y(w) \land \neg t(w) \land \neg e(w) \\
& \land & r_x(w) \land r_y(w) \land \neg r_t(w) \land \neg r_e(w) \land \neg is(w)\\
%% formula
\hline
\characteristicFormula{S_2}& = & \Land_{i=1,2} (\exists v: \nodeFormula{S_2}{u_i}(v)) \\
    & \land &\forall w: \Lor_{i=1,2} \nodeFormula{S_2}{u_i}(w)\\
    & \land & \forall w_1,  w_2: \Land_{i=1,2} \nodeFormula{S_1}{u_i}(w_i) \implies \\
    && \neg eq(w_1, w_2) \land \neg n(w_2, w_1) \land n(w1, w2)\\
    & \land & \forall w_1,  w_2: \Land_{i=1,2} \nodeFormula{S_1}{u_1}(w_i) \implies \\
    && \land eq(w_1, w_2) \land \neg n(w_1, w_2) \\
    & \land & \forall w_1,  w_2: \Land_{i=1,2} \nodeFormula{S_1}{u_2}(w_i) \implies \\
    && \land eq(w_1, w_2) \land \neg n(w_1, w_2) \\
\end{array}
\\
  \hline
  \xymatrix@R10pt@C10pt{
   \tx\ar[r]          &  
        \xyNonSummaryNode{u_1}\ar[r]^n &
        \xyNonSummaryNode{u_2}\ar@{.>}[r]^n &
        \xySummaryNode{u_3}\ar@{.>}@(ur, ul)[]|{n}\\
S_3	&r_x \ar[u] & y, r_x, r_y \ar[u] & r_x, r_y \ar[u] 
}

   &
\begin{array}{llll}
%% S_3
%%% nodes
%%
\nodeFormula{S_3}{u_1}(w)&=&& x(w) \land \neg y(w) \land \neg t(w) \land \neg e(w)\\
& \land && r_x(w) \land \neg r_y(w) \land \neg r_t(w) \land \neg r_e(w) \land \neg is(w)\\
\nodeFormula{S_3}{u_2}(w) & =&& \neg x(w) \land y(w) \land \neg t(w) \land \neg e(w) \\
& \land && r_x(w) \land r_y(w) \land \neg r_t(w) \land \neg r_e(w) \land \neg is(w)\\
\nodeFormula{S_3}{u_3}(w) & =&& \neg x(w) \land \neg y(w) \land \neg t(w) \land \neg e(w) \\
& \land && r_x(w) \land r_y(w) \land \neg r_t(w) \land \neg r_e(w) \land \neg is(w)\\
%% formula
\hline
\characteristicFormula{S_3}& = && \Land_{i=1,2,3} (\exists v: \nodeFormula{S_3}{u_i}(v))\\
    & \land && \forall w: \Lor_{i=1,2,3} \nodeFormula{S_3}{u_i}(w)\\
    & \land && \forall w_1,  w_2: (\Land_{i=1,2} \nodeFormula{S_3}{u_1}(w_i) \implies \\
    &&& eq(w_1, w_2) \land \neg n(w_1, w_2)) \\
    && \land & (\Land_{i=1,2} \nodeFormula{S_3}{u_2}(w_i) \implies \\
    &&& eq(w_1, w_2) \land \neg n(w_1, w_2)) \\
    && \land & (\nodeFormula{S_3}{u_1}(w_1) \land \nodeFormula{S_3}{u_2}(w_2) \implies \\
    &&& \neg eq(w_1, w_2) \land \neg n(w_2, w_1) \land n(w_1, w_2))\\
    && \land & (\nodeFormula{S_3}{u_2}(w_1) \land \nodeFormula{S_3}{u_3}(w_2) \implies \\
    &&& \neg eq(w_1, w_2) \land \neg n(w_2, w_1)) \\
    && \land & (\nodeFormula{S_3}{u_1}(w_1) \land \nodeFormula{S_3}{u_3}(w_2) \implies \\
    &&& \neg eq(w_1, w_2) \land \neg n(w_2, w_1) \land \neg n(w_1, w_2)) \\
\end{array}
\\
  \hline
\end{array}$
}
\caption{\label{Ta:StructFormula} (Continued in \tableref{StructFormulaTwo}.) The left column shows
the structures that arise at the beginning of the loop in the \tinsert\/ program
from \figref{Insert}(b).
The right column shows the characteristic formula for each structure.
Note that we omit the redundant sub-formulas $\characteristicFormula{S}[p]$,
for $p \in \Voc_1$, that are part of
$\characteristicFormula{S}_{total}$ and $\nodeFormula{S_i}{u_j}(w)$
definitions.}
\end{table}

\begin{table}
\centering
\resizebox{\textwidth}{!}{
$\begin{array}{|l|l|}
\hline
{\bf Structure\/} & {\bf Characteristic Formula\/}\\
\hline
  \xymatrix@R10pt@C10pt{
   \tx\ar[r]          &
        \xyNonSummaryNode{u_1}\ar@{.>}[r]^n &
        \xySummaryNode{u_2}\ar@{.>}[r]^n\ar@{.>}@(ur, ul)[]|{n} &
        \xyNonSummaryNode{u_3}\ar@{.>}[r]^n &
        \xySummaryNode{u_4}\ar@{.>}@(ur, ul)[]|{n}\\
   S_4  &r_x\ar[u] & r_x\ar[u] & y, r_x, r_y \ar[u] & r_x, r_y \ar[u]
}
 &
  \begin{array}{llll}
%% S_3
%%% nodes
%%
\nodeFormula{S_4}{u_1}(w)&=&& x(w) \land \neg y(w) \land \neg t(w) \land \neg e(w)\\
& \land && r_x(w) \land \neg r_y(w) \land \neg r_t(w) \land \neg r_e(w) \land \neg is(w)\\
\nodeFormula{S_4}{u_1}(w)&=&& \neg x(w) \land \neg y(w) \land \neg t(w) \land \neg e(w)\\
& \land && r_x(w) \land \neg r_y(w) \land \neg r_t(w) \land \neg r_e(w) \land \neg is(w)\\
\nodeFormula{S_4}{u_3}(w) & =&& \neg x(w) \land y(w) \land \neg t(w) \land \neg e(w) \\
& \land && r_x(w) \land r_y(w) \land \neg r_t(w) \land \neg r_e(w) \land \neg is(w)\\
\nodeFormula{S_4}{u_4}(w) & =&& \neg x(w) \land \neg y(w) \land \neg t(w) \land \neg e(w) \\
& \land && r_x(w) \land r_y(w) \land \neg r_t(w) \land \neg r_e(w) \land \neg is(w)\\
%% formula
\hline
\characteristicFormula{S_1}& = &&
    \Land_{i=1,\ldots,4} (\exists v: \nodeFormula{S_4}{u_i}(v))\\
    & \land && \forall w: \Lor_{i=1,\ldots,4} \nodeFormula{S_4}{u_i}(w)\\
    & \land & \forall w_1,  w_2:& \\
    &&& (\Land_{i=1,2} \nodeFormula{S_4}{u_1}(w_i) \implies \\
    &&& eq(w_1, w_2) \land \neg n(w_1, w_2)) \\
    && \land & (\Land_{i=1,2} \nodeFormula{S_4}{u_3}(w_i) \implies \\
    &&& eq(w_1, w_2) \land \neg n(w_1, w_2)) \\
    && \land & (\nodeFormula{S_4}{u_1}(w_1) \land \nodeFormula{S_4}{u_2}(w_2) \implies \\
    &&& \neg eq(w_1, w_2) \land \neg n(w_2, w_1) )\\
    && \land & (\nodeFormula{S_4}{u_2}(w_1) \land \nodeFormula{S_4}{u_3}(w_2) \implies \\
    &&& \neg eq(w_1, w_2) \land \neg n(w_2, w_1)) \\
    && \land & (\nodeFormula{S_4}{u_1}(w_1) \land \nodeFormula{S_4}{u_3}(w_2) \implies \\
    &&& \neg eq(w_1, w_2) \land \neg n(w_2, w_1) \land \neg n(w_1, w_2)) \\
    && \land & (\nodeFormula{S_4}{u_3}(w_1) \land \nodeFormula{S_4}{u_4}(w_2) \implies \\
    &&& \neg eq(w_1, w_2) \land \neg n(w_2, w_1)) \\
    && \land & (\nodeFormula{S_4}{u_1}(w_1) \land \nodeFormula{S_4}{u_4}(w_2) \implies \\
    &&& \neg eq(w_1, w_2) \land \neg n(w_2, w_1) \land \neg n(w_1, w_2)) \\
    && \land & (\nodeFormula{S_4}{u_2}(w_1) \land \nodeFormula{S_4}{u_4}(w_2) \implies \\
    &&& \neg eq(w_1, w_2) \land \neg n(w_2, w_1) \land \neg n(w_1, w_2)) \\
\end{array}
\\
  \hline
  \xymatrix@R10pt@C10pt{
   \tx\ar[r]          &
        \xyNonSummaryNode{u_1}\ar@{.>}[r]^n &
        \xySummaryNode{u_2}\ar@{.>}[r]^n\ar@{.>}@(ur, ul)[]|{n} &
        \xyNonSummaryNode{u_3}\\
S_5 &r_x \ar[u] & r_x \ar[u] & y, r_x, r_y \ar[u]
}
 &
  \begin{array}{llll}
%% S_3
%%% nodes
%%
\nodeFormula{S_5}{u_1}(w)&=&& x(w) \land \neg y(w) \land \neg t(w) \land \neg e(w)\\
& \land && r_x(w) \land \neg r_y(w) \land \neg r_t(w) \land \neg r_e(w) \land \neg is(w)\\
\nodeFormula{S_5}{u_2}(w) & =&& \neg x(w) \land \neg y(w) \land \neg t(w) \land \neg e(w) \\
& \land && r_x(w) \land \neg r_y(w) \land \neg r_t(w) \land \neg r_e(w) \land \neg is(w)\\
\nodeFormula{S_5}{u_3}(w) & =&& \neg x(w) \land y(w) \land \neg t(w) \land \neg e(w) \\
& \land && r_x(w) \land r_y(w) \land \neg r_t(w) \land \neg r_e(w) \land \neg is(w)\\
%% formula
\hline
\characteristicFormula{S_3}& = && \Land_{i=1,2,3}(\exists v: \nodeFormula{S_5}{u_i}(v)) \\
    & \land && \forall w: \Lor_{i=1,2,3} \nodeFormula{S_5}{u_i}(w)\\
    & \land & \forall w_1,  w_2: & \\
    &&& (\Land_{i=1,2} \nodeFormula{S_5}{u_1}(w_i) \implies \\
    &&& eq(w_1, w_2) \land \neg n(w_1, w_2)) \\
    && \land & (\Land_{i=1,2} \nodeFormula{S_5}{u_3}(w_i) \implies \\
    &&& eq(w_1, w_2) \land \neg n(w_1, w_2)) \\
    && \land & (\nodeFormula{S_5}{u_1}(w_1) \land \nodeFormula{S_5}{u_2}(w_2) \implies \\
    &&& \neg eq(w_1, w_2) \land \neg n(w_2, w_1))\\
    && \land & (\nodeFormula{S_5}{u_2}(w_2) \land \nodeFormula{S_5}{u_3}(w_2) \implies \\
    &&& \neg eq(w_1, w_2) \land \neg n(w_2, w_1)) \\
    && \land & (\nodeFormula{S_5}{u_1}(w_1) \land \nodeFormula{S_5}{u_3}(w_2) \implies \\
    &&& \neg eq(w_1, w_2) \land \neg n(w_2, w_1) \land \neg n(w_1, w_2)) \\
\end{array}
\\
\hline
\end{array}$
}
\caption{\label{Ta:StructFormulaTwo}\tableref{StructFormula} continued.}
\end{table}

\section{Proofs}\label{Se:Proofs}
\newtheorem{SubLemma}{Lemma}[section]

%________________________________________________________________________
\begin{SubLemma}\label{Lem:embedding3color}
Consider the $3$-valued structure $S$ shown in \figref{Color}.
For all $2$-valued structures $C$,
$C$ can be embedded into $S$ if and only if $C$ can be colored using $3$ colors.
\begin{ProofIf} Suppose that $C$ is 3-colorable, let $c$ be a mapping from the
nodes of $C$ to the colors $\{1,2,3\}$.
We define embedding function $f$ from $C$ to $S$ as follows:
$f(u) = u_{c(u)}$, i.e., a node $u \in C$ that has color $i$ is mapped to $u_i \in S$.
It is easy to see that $f$ preserves predicate values in $S$,
because the only definite values in $S$ indicate the absence of self-loops.
It is preserved,
because there are no edges in $C$ with both endpoints in the same color.
\end{ProofIf}
\begin{ProofOnlyIf}
Suppose that $C$ is embedded into $S$ using $f$.
We show that $C$ is 3-colorable. For each node $u \in C$, let the color of u, $c(u)$, be the  name of the corresponding node in $S$,
i.e., $c(u) = f(u)$.
The absence of self loops on any of the three summary nodes guarantees that a pair of adjacent
nodes in $C$ cannot be mapped by $f$ to the same summary node.
That is, for any edge in $C$ the endpoints must be mapped by $f$ to different summary nodes,
thus they have different colors.
\end{ProofOnlyIf}
\end{SubLemma}
%________________________________________________________________________
\begin{SLem}{unique}
Let $S$ be an FO-identifiable structure and let $u_1, u_2 \in S$ be distinct individuals.
Let $\C{S}$ be a 2-valued structure that embeds into $S$ and let $\C{u} \in \C{S}$.
At most one of the following can hold, but not both:
\begin{enumerate}
\item $\C{S},[w \mapsto \C{u}] \models \nodeFormula{S}{u_1}(w)$
\item $\C{S},[w \mapsto \C{u}] \models \nodeFormula{S}{u_2}(w)$
\end{enumerate}
\begin{proof}
Because $\C{S}$ embeds into $S$, there exists an embedding function $f$,
such that $\C{S} \sqsubseteq^f S$.
For the sake of argument, assume that both claims hold.
By \defref{UniqueFONames}, we get that $f(\C{u}) = u_1$ and $f(\C{u}) = u_2$;
because $f$ is a function, we get that $u_1 = u_2$.
This yields a contradiction to the assumption that $u_1$ and $u_2$ are distinct individuals.
\end{proof}
\end{SLem}
%________________________________________________________________________
\begin{SLem}{Trivial}
For every $2$-valued structure $\C{S}$ and assignment $Z$
\[
\C{S}, Z \models p^B(v_1, v_2, \ldots, v_k) ~\mbox{iff}~
\iota^{\C{S}}(p)(Z(v_1), Z(v_2), \ldots, Z(v_k)) \sqsubseteq B
\]
\begin{ProofIf}
Suppose that $\iota^{\C{S}}(p)(Z(v_1), Z(v_2), \ldots, Z(v_k))
\sqsubseteq B$. There are two cases to consider: (i)~$B=1/2$ or
(ii)~$\iota^{\C{S}}(p)(Z(v_1), Z(v_2), \ldots, Z(v_k)) = B$. If
$B=1/2$, then by \defref{pb}, $p^B(v_1, v_2, \ldots, v_k)=1$ and
thus $\C{S}, Z \models p^B(v_1, v_2, \ldots, v_k)$
for all $Z$.
If B = 1, then $\iota^{\C{S}}(p)(Z(v_1), Z(v_2), \ldots, Z(v_k)) = 1$,
thus $\C{S}, Z \models
p(v_1, v_2, \ldots, v_k)$ which is $\C{S}, Z \models p^1(v_1, v_2,
\ldots, v_k)$ by \defref{pb}. Similarly, if B = 0, then
$\iota^{\C{S}}(p)(Z(v_1), Z(v_2), \ldots, Z(v_k)) = 0$ implies
that $\C{S}, Z \models \neg p(v_1, v_2, \ldots, v_k) = p^0(v_1,
v_2, \ldots, v_k)$.
\end{ProofIf}
\begin{ProofOnlyIf}
Assume that $\C{S}, Z \models p^B(v_1, v_2, \ldots, v_k)$. If
$B=1/2$, then
%\[
$\iota^{\C{S}}(p)(Z(v_1), Z(v_2), \ldots, Z(v_k)) \sqsubseteq B$
%\]
trivially holds. If $B=0$, apply \defref{pb} to the assumption to
get $\C{S}, Z \models \neg p(v_1, v_2, \ldots, v_k)$, which
implies \newline
$\iota^{\C{S}}(p)(Z(v_1), Z(v_2), \ldots, Z(v_k)) = 0 =
B$. Similarly, if $B=1$, the
assumption implies $\iota^{\C{S}}(p)(Z(v_1), Z(v_2), \ldots,
Z(v_k)) = 1 = B$.
\end{ProofOnlyIf}
\end{SLem}
%________________________________________________________________________
\begin{SLem}{Bounded}
Every bounded $3$-valued structure $S$ is FO-identifiable, where
\[
\nodeFormula{S}{u_i}(w)  \eqdef \Land_{p \in \Voc_1}
p^{\iota^{S}(p)(u_i)}(w)
\]
\begin{Proof}
Consider a bounded $3$-valued structure $S = \{U, \iota^S\}$.
We shall show that every element
$u \in U$ is FO-identifiable using the formula defined in \equref{individualFormula}.
Let $\C{S}$ be a $2$-valued structure
that embeds into $S$ using a function $f$,
and let $\C{u}$ be a concrete element in $U^{\C{S}}$.
By \defref{UniqueFONames}, we have to show that the following holds:
\[
f(\C{u})= u
\iff
\C{S},[w \mapsto \C{u}] \models \nodeFormula{S}{u}(w)
\]
\begin{ProofIf}
Suppose that $\C{S},[w \mapsto \C{u}] \models \nodeFormula{S}{u}(w)$.
In particular, each conjunct of $\nodeFormula{S}{u}$ must hold, i.e.,
for each predicate $p \in \Voc_1$,
$\C{S},[w \mapsto \C{u}] \models p^{\iota^{S}(p)(u)}(w)$.
Using \lemref{Trivial} we get that $\iota^{\C{S}}(p)(\C{u}) \sqsubseteq \iota^{S}(p)(u)$.
In addition, the embedding condition in \equref{EmbeddingCondition},
requires, in particular, that for each unary predicate $p$
$\iota^{\C{S}}(p)(\C{u}) \sqsubseteq \iota^{S}(p)(f(\C{u}))$ holds.
Let $u_1 = f(\C{u})$.
For the sake of argument, assume that $u_1 \neq u$.
Recall that $S$ is a bounded structure,
in which every individual must have a unique
combination of definite values of unary predicates.
As a consequence, there must be a unary predicate $p$
such that $\iota^S(p)(u_1) \neq \iota^S(p)(u)$ and
the value of $p$ on both $u_1$ and $u$ is definite.
This yields a contradiction,
because $\sqsubseteq$ on definite values
implies equality; however
$\iota^{\C{S}}(p)(\C{u}) = \iota^{S}(p)(u)$
and
$\iota^{\C{S}}(p)(\C{u}) = \iota^{S}(p)(f(\C{u})) = \iota^{S}(p)(u_1)$
can not hold simultaneously, by the assumption.
\end{ProofIf}
\begin{ProofOnlyIf}
Suppose that $f(\C{u}) = u$.
Using \equref{EmbeddingCondition},
the embedding function $f$ guarantees that for each unary predicate $p$,
$\iota^{\C{S}}(p)(\C{u}) \sqsubseteq \iota^{S}(p)(f(\C{u}))$.
This means that
$\C{S},[w \mapsto \C{u}] \models p^{\iota^{S}(p)(f(\C{u}))}(w)$
by \lemref{Trivial}, or
$\C{S},[w \mapsto \C{u}] \models p^{\iota^{S}(p)(u)}(w)$
by the assumption.
This holds for all unary predicates, and thus
holds for their conjunction as well, namely, for
the formula $\nodeFormula{S}{u}$.
\end{ProofOnlyIf}
\end{Proof}
\end{SLem}
%________________________________________________________________________
\begin{SubLemma}\label{Lem:focusFOIdent}
Given a set of formulas $F$ and a $3$-valued structure $S$,
if the ``focus'' algorithm~\cite[Sec.6]{Thesis:LevAmi00} terminates,
it returns a set of structures $X$ such that
$\gamma(S) = \gamma(X)$ and every formula $\varphi \in F$ evaluates, using the compositional semantics,
to a definite value in every structure
in $X$, for every assignment.
If the input structure $S$ is FO-Identifiable, then all structures in $X$ are FO-Identifiable.
\begin{Proof}
By induction on the iterations of the loop in the ``focus'' algorithm,
it is sufficient to show that the structures returned by the procedure {\tt FocusAssignment}
from \cite[Fig.17]{Thesis:LevAmi00} are FO-Identifiable.
The only interesting case is when the input literal of {\tt FocusAssignment} is of the form $p(u_1, \ldots, u_k)$.
The resulting set of structures $X$ is $\{ S_0, S_1, S'' \}$ where $S_0$ and $S_1$ are copies
of $S$ with  $p(u_1, \ldots, u_k)$ set to $0$ and $1$, respectively.
Thus, if $S$ is FO-identifiable, then $S_0$ and $S_1$ are FO-identifiable.
$S''$ is a result of splitting a node $u_i \in S$ into $u.0$ and $u.1$,
and setting $p(u_1, \ldots, u_k)$ to $0$ on one of the copies, and to $1$ on the other.
To simplify the exposition, suppose that the first node $u_1$ is split.
Then $S''$ is FO-identifiable using the formulas $\nodeFormula{S}{u}(w)$ for all $u$ except $u.0, u.1$, and
\[
\begin{array}{l}
\nodeFormula{S''}{u.0}(w) \eqdef \exists v_2, \ldots, v_k . \neg p(w, v_2, \ldots, v_k) \land \nodeFormula{S}{u}(w) \land
\Land_{j=2,\ldots, k} \nodeFormula{S}{u_j}(v_j)\\
\nodeFormula{S''}{u.1}(w) \eqdef \exists v_2, \ldots, v_k . p(w, v_2, \ldots, v_k) \land \nodeFormula{S}{u}(w) \land
\Land_{j=2,\ldots, k} \nodeFormula{S}{u_j}(v_j)
\end{array}
\]
\end{Proof}
\end{SubLemma}
%________________________________________________________________________
\begin{SThe}{RepresentingStructuresByFormulae}
For every FO-identifiable $3$-valued structure $S$,
and $2$-valued structure $\C{S}$
\[
\C{S} \in \gamma(S)~ \mbox{iff}~\C{S} \models
\widehat{\gamma}(S)
\]
\begin{Proof}
In \lemref{RepresentIf}, we show that the if-direction holds,
even when $S$ is not FO-identifiable, i.e.,
every concrete structure satisfying the
characteristic formula $\widehat{\gamma}(S)$ is indeed in $\gamma(S)$.
In \lemref{RepresentOnlyIf} we show the only-if part, i.e.,
for an FO-identifiable structure, the other direction is also true.
\end{Proof}
\end{SThe}
%________________________________________________________________________
\begin{SubLemma}\label{Lem:RepresentIf}
Let $S$ be a first-order structure with set of individuals
$U=\{u_1, u_2, \ldots, u_n\}$.
Let $\nodeFormula{S}{u_i}(w)$ used in $\widehat{\gamma}(S)$ be
an arbitrary first-order formula free in $w$,
such that \lemref{unique} holds.
Then, for all $\C{S}$ such that
$\C{S} \models \widehat{\gamma}(S)$, $\C{S} \in \gamma(S)$.
\begin{Proof}
Let $\C{S}=\B{\C{U}, \C{\iota}}$ be a concrete structure such that $\C{S} \models
\widehat{\gamma}(S)$.
We shall construct a surjective function $f\colon \C{U} \to U$ such that
$\C{S} \sqsubseteq^f S$.
Let $\C{Z}$ be an assignment over $v_1, \ldots, v_n$ such that $\C{S}, \C{Z}
\models \varphi$, where $\varphi \eqdef \Land_{i=1}^n \nodeFormula{S}{u_i}(v_i)$,
i.e., $\varphi$ is the first line of \equref{characteristicFormula} without the existential
quantification.
Note that all $\C{Z}(v_i)$ are distinct, according to \lemref{unique}.
Define the function $f\colon \C{U} \to U$ by:
\begin{equation}
f(\C{u}) = \left \{ \begin{array}{ll}
    u_i & \mbox{if}~\C{Z}(v_i) = \C{u}\\
    u_j & \mbox{if for all }i,~\C{Z}(v_i) \neq \C{u}~\mbox{and}~u_j~\mbox{is an arbitrary element such that}\\
  & \C{S}, [w \mapsto \Conc{u}] \models \nodeFormula{S}{u_j}(w)
        \end{array}
        \right .
\label{eq:CandidateEmbedding}
\end{equation}

Let us show that every concrete element is mapped to some element
in $U$. In the case that $Z(v_i)= \C{u}$, the concrete element $\C{u}$
is mapped to $u_i \in U$ by $f$.
Otherwise, because $\C{S} \models \characteristicFormula{S}[total]$
holds, at least one of its disjuncts must be satisfied by each $\C{u}$,
i.e. $\C{S}, [w \mapsto \C{u}]$ must satisfy $\nodeFormula{S}{u_j}(w)$ for
some $u_j$; thus $f$'s definition will map $\C{u}$ to this $u_j$.
Therefore, $f(\C{u})$ is well-defined.

In addition, every element $u_i \in U$ is assigned by $f$ to some concrete
element $\C{u}_i \in \C{U}$ such that $Z(v_i)= \C{u}_i$.
%Since $\C{S}, \C{Z}$ satisfies the sub-formula $\Land_{k \neq j} \neg
%eq(v_k, v_j)$,
According to \lemref{unique}, all such elements $\C{u}_i$ are different.
Therefore, $f(\C{u})$ is surjective.

Let $p$ be a nullary predicate.
Because $\C{S}$ satisfies $\characteristicFormula{S}_{nullary}$,
it must satisfy each conjunct, in particular
$\C{S} \models p^{\iota^{S}(p)()}$.
Using \lemref{Trivial} we get that
$\iota^{\C{S}}(p)() \sqsubseteq \iota^{S}(p)()$.

Let $p \in P$ be a predicate of arity $r \geq 1$.
Let $\C{u}_1, \C{u}_2, \ldots, \C{u}_r \in \C{U}$ and let us show that
\begin{equation}\label{eq:naryProof}
\iota^{\C{S}}(p)(\C{u}_1, \C{u}_2, \ldots, \C{u}_r) \sqsubseteq
\iota^S(p)(f(\C{u}_1), f(\C{u}_2), \ldots, f(\C{u}_r))
\end{equation}
Let $Z$ be an assignment such that $Z(w_i) = \C{u}_i$ for\ $i= 1, \ldots, r$.
Because $\C{S} \models \characteristicFormula{S}[p]$,
we conclude that
$\C{S}, Z$ satisfies the body of \equref{naryFormula}.
Consider the conjunct of the body with premise
$\Land_{j=1}^r \nodeFormula{S}{f(\C{u}_j)}(w_j)$.
%If the premise does not hold, it imposes no limitations on $p$'s interpretation
%under $S$ and \equref{naryProof} holds.
By definition of $f$, $\C{S}, w_j \mapsto \C{u}_j$ satisfies $\nodeFormula{S}{f(\C{u}_j)}(w_j)$
for all $j=1, \ldots, r$,
which means that the premise is satisfied by\ $\C{S}, Z$.
Therefore, the conclusion must hold:
%Otherwise, if the premise holds for\ $\C{S}, Z$, then the conclusion must hold:
$\C{S}, Z \models p^{\iota^S(p)(f(\C{u}_{1}), \ldots, f(\C{u}_{r}))}(w_1, \ldots, w_r))$
and the result follows from \lemref{Trivial}.
\end{Proof}
\end{SubLemma}
%________________________________________________________________________

\begin{SubLemma}\label{Lem:RepresentOnlyIf}
For every $3$-valued FO-identifiable structure $S$,
and $2$-valued structure $\C{S}$ such that $\C{S} \models F$ and
$\C{S} \sqsubseteq S$, $\C{S} \models \characteristicFormula{S}$.
\begin{Proof}
Let $f\colon \C{S} \to S$ be a surjective function such that
$\C{S} \sqsubseteq^f S$.
Let $\C{u}_i$ be an arbitrary element such that $f(\C{u}_i) = u_i$.
Define an assignment $\C{Z}$ such that $\C{Z}(v_i) = \C{u}_i$;
$\C{u}_i$ must exist because $f$ is surjective.
Because $S$ is FO-identifiable, by \defref{UniqueFONames}
we conclude that for every $1 \leq i \leq n$,
$\C{S}, \C{Z} \models \nodeFormula{S}{u_i}(v_i)$.
Because $f$ is a function, all $\C{u}_i$ are distinct elements,
according to \lemref{unique}.
%meaning that $\C{S}, \C{Z}$ satisfies the sub-formula $\Land_{k \neq j} \neg eq(v_k, v_j)$.

Because $f$ is a function, for every $\C{u}$ there is $u$ such that $f(\C{u}) = u$.
Then, by \defref{UniqueFONames}, $\C{S}, [w \mapsto \C{u}] \models \nodeFormula{S}{u}(w)$,
i.e., every assignment to $w$ in $\C{S}$ satisfies some disjunct of $\characteristicFormula{S}_{total}$.
That is $\C{S}$ satisfies $\characteristicFormula{S}_{total}$.

For every nullary predicate $p \in \Voc_0$, using \equref{EmbeddingCondition}
and \lemref{Trivial}, we conclude that $\C{S}$ satisfies $p^{\iota^{S}(p)()}$.
Therefore, $\C{S}$ satisfies $\characteristicFormula{S}_{nullary}$.

Let $p \in P$ be a predicate of arity $r$.
Let $\C{u}_1, \ldots, \C{u}_r \in \C{U}$ and let $\C{Z}$ be an assignment
such that $\C{Z}(w_i) = \C{u}_i$.
We shall show that $\C{S}, \C{Z}$ satisfy the body of \equref{naryFormula}.
If the premise of the implication is not satisfied then the formula vacuously holds.
Otherwise, $\C{S}, \C{Z} \models \nodeFormula{S}{u_i}(w_i)$ for all $i = 1, \ldots, r$.
Then, by \defref{UniqueFONames}, $f(\C{u}_i) = u_i$.
Using \equref{EmbeddingCondition} on $f$, we get
$\iota^{\C{S}}(p)(\C{u}_1, \ldots, \C{u}_r) \sqsubseteq \iota^S(p)(f(\C{u}_1), \ldots, f(\C{u}_r))$,
which means that $\iota^{\C{S}}(p)(\C{u}_1, \ldots, \C{u}_r) \sqsubseteq \iota^S(p)(u_1, \ldots, u_r)$ holds.
By \lemref{Trivial}, we conclude that
$\C{S},\C{Z}$ satisfies $p^{\iota^S(p)(u_1, \ldots, u_r)}(w_1, \ldots, w_r)$.
\end{Proof}
\end{SubLemma}
%________________________________________________________________________

\begin{SLem}{ICAProp}
If\ $3$-valued structure $S = \B{U, \iota^S}$ over vocabulary $\Voc$
is \ICA\, then:
\begin{description}
    \item [(i)] $S$ is a bounded structure.
    \item [(ii)] For each nullary predicate $p$, $\iota^S(p)() \in \{0,1\}$.
    \item [(iii)] For each element $u \in U$, and each unary predicate $p$, $\iota^S(p)(u) \in \{0,1\}$.
\end{description}
\begin{Proof}
%Let $S = \B{U, \iota^S}$ be an \ICA\ of a $2$-valued structure $\C{S}$.
Let $\C{S} = \{\C{U}, \iota^{\C{S}}\}$ be a $2$-valued structure, such that $S$ is the canonical
abstraction of $\C{S}$.
Let $\blur \colon \C{U} \to U$ be the mapping that identifies $S$ as the canonical abstraction of $\C{S}$.
\begin{description}
\item [(i)] Show that $S$ is a bounded structure.
By \equref{tightEmbed}, every abstract element represents concrete
elements with the same canonical name.
Thus, for two distinct abstract elements $u_0, u_1 \in U^S$,
the canonical name of concrete elements
represented by $u_0$ is different from the canonical name of concrete
elements represented by $u_1$.
Without loss of generality, assume that the canonical names differ in
a unary predicate $p$, such that $p$ evaluates to $0$ on all concrete elements
represented by $u_0$, and $p$ evaluates to $1$ on all concrete elements represented by $u_1$.
From the join operation in \equref{canonicName}, it follows that the value of $p$ on $u_0$ must be $0$
and the value of $p$ on $u_1$ must be $1$.
This shows that, in general, every pair of distinct elements in $S$
differs in a definite value of some unary predicate,
proving that $S$ is a bounded structure.
\item [(ii)] Let $p$ be a nullary predicate. Show that $\iota^S(p)() \in \{0,1\}$.
By \equref{canonicName},
$\iota^S(p)() = \sqcup \{\iota^{\C{S}}(p)() \} = \iota^{\C{S}}(p)()$.
This means that $p$ has the same value in $S$ and $\C{S}$.
Because $\C{S}$ is a concrete structure, the value of $p$ must be definite.
\item [(iii)] Let $p$ be a unary predicate and let $u \in U$.
Show that $\iota^S(p)(u) \in \{0,1\}$.
Suppose that the opposite holds: $\iota^S(p)(u) = 1/2$.
By \equref{canonicName}, there exist two concrete elements,
denoted by $u_0$ and $u_1$,
such that $\blur(u_0) = u$ and $\blur(u_0)=u$, and
$p$ evaluates to $0$ on $u_0$ and to $1$ on $u_1$.
Hence, these concrete elements have different canonical names
and by \equref{tightEmbed} they cannot be mapped by $\blur$ to
the same abstract element; this contradicts the supposition and hence
$\iota^S(p)(u) \in \{0,1\}$.
\end{description}
\end{Proof}
\end{SLem}
%________________________________________________________________________
\begin{SLem}{CanonicBounded}
Every $3$-valued structure $S$ that is an \ICA\, is canonical-FO-identifiable, where
\begin{equation}
\nodeFormula{S}{u_i}(w)  \eqdef \Land_{p \in \Voc_1} %AbstractionPredicates}
p^{\iota^{S}(p)(u_i)}(w)
\end{equation}
\begin{Proof}
Let $S = \{U, \iota^S\}$ be a $3$-valued structure that is \ICA.
We shall show that every element $u \in U$ is
canonical-FO-identifiable using the formula defined in \equref{canonIndividualFormula}.
Let $\C{S}=\{\C{U}, \iota^{\C{S}}\}$
be a $2$-valued structure, such that $S$ is the
canonical abstraction of $\C{S}$, induced by a function $\blur$,
and let $\C{u} \in U^{\C{S}}$.
By \defref{CanonicUniqueFONames}, we have to show that the following holds:
\[
\blur(\C{u})= u
\iff
\C{S},[w \mapsto \C{u}] \models \nodeFormula{S}{u}(w)
\]
\begin{ProofIf}
Suppose that $\C{S},[w \mapsto \C{u}] \models \nodeFormula{S}{u}(w)$.
Let $u_1 = \blur(\C{u})$.
For the sake of argument, assume that $u_1 \neq u$.
$S$ is an \ICA\ and using \lemref{ICAProp}(i) we get that $S$ is a bounded structure.
By \defref{BoundedStructures}, there exists a unary predicate $p$
that evaluates to different definite values on $u$ and $u_1$.
Without loss of generality, suppose that $p$ evaluates to $0$ on $u$ and to $1$ on $u_1$.
This implies the following two facts.
First, from property \equref{canonicName} of the definition of canonical abstraction,
$p$ also evaluates to $1$ on all concrete values mapped to $u_1$ by $\blur$;
in particular, $p$ must evaluate to $1$ on $\C{u}$.
Second, recall that by assumption, each conjunct of $\nodeFormula{S}{u}$ must hold, i.e.,
for each predicate $p \in \Voc_1$, $\C{S},[w \mapsto \C{u}] \models p^{\iota^{S}(p)(u)}(w)$.
Because $p$ evaluates to $0$ on $u$,
we get from \defref{pb} that $\C{S},[w \mapsto \C{u}] \models p^0(w)$,
which means $\iota^{\C{S}}(p)(\C{u}) = 0$ and a contradiction is obtained.
\end{ProofIf}
\begin{ProofOnlyIf}
Suppose that $\blur(\C{u}) = u$.
Because $S$ is an \ICA\, by \lemref{ICAProp}(iii) we know that
all unary predicates have definite values in $S$.
Let $p$ be a unary predicate. Let $B \in \{1, 0\}$ be such that $\iota^S(p)(u) = B$.
Because $p$ has definite value $B$ on $u$ in $S$, by \equref{canonicName}
it must have the same definite value $B$ on all concrete nodes in $\C{S}$
that are mapped to $u$ by $\blur$;
in particular, on $\C{u}$: $\iota^{\C{S}}(p)(\C{u}) = B$.
Therefore, using \defref{pb}, $\C{S}, [w \mapsto \C{u}] \models p^B(w)$,
in other words, $\C{S},[w \mapsto \C{u}] \models p^{\iota^{S}(p)(u)}(w)$.
This holds for all unary predicates, and thus
holds for their conjunction as well, i.e., for
the formula $\nodeFormula{S}{u}$.
\end{ProofOnlyIf}
\end{Proof}
\end{SLem}
%________________________________________________________________________
\begin{SThe}{RepresentingTightEmbeddingByFormulae}
For every $3$-valued structure $S$ that is an \ICA\,
and $2$-valued structure $\C{S}$
\[
\C{S} \in \gamma_c(S)
~\mbox{iff}~\C{S} \models \widehat{\gamma}_c(S)
\]
\begin{Proof}
In \lemref{CanonicRepresentIf}, we show that the if-direction holds, i.e.,
a $3$-valued structure $S$ is the canonical abstraction of every concrete structure satisfying the
characteristic formula $\widehat{\gamma_c}(S)$;
in \lemref{CanonicRepresentOnlyIf} we show the other direction.
\end{Proof}
\end{SThe}
%________________________________________________________________________
\begin{SubLemma}\label{Lem:CanonicRepresentIf}
Let $S$ be an \ICA\ with set of individuals
$U=\{u_1, u_2, \ldots, u_n\}$.
Let $\nodeFormula{S}{u_i}(w)$ be an arbitrary formula free in $w$, used in $\widehat{\gamma_c}$,
such that \lemref{unique} holds.
Then, for all $\C{S}$ such that
$\C{S} \models \widehat{\gamma_c}(S)$, $S$ is a canonical abstraction of $\C{S}$.
\begin{Proof}
Let $\C{S}=\B{\C{U}, \C{\iota}}$ be a concrete structure
such that $\C{S} \models \widehat{\gamma_c}(S)$.
We shall construct a surjective function $\blur\colon \C{U} \to U$ such that
$\C{S}$ is a canonical abstraction of $S$.
From \defref{TightChar} it follows, in particular, that $\C{S} \models \characteristicFormula{S}$.
Let $\C{Z}$ be an assignment over $v_1, \ldots, v_n$ such that $\C{S}, \C{Z}
\models \varphi$, where $\varphi \eqdef \Land_{i=1}^n \nodeFormula{S}{u_i}(v_i)$,
i.e., $\varphi$ is the first line of
\equref{characteristicFormula} without the existential
quantification).
Note that all $\C{Z}(v_i)$ are distinct, according to \lemref{unique}.
Define the function $\blur\colon \C{U} \to U$ by:
\begin{equation}
\blur(\C{u}) = \left \{ \begin{array}{ll}
    u_i & \mbox{if}~\C{Z}(v_i) = \C{u}\\
    u_j & \mbox{if for all }i,~\C{Z}(v_i) \neq \C{u}~\mbox{and}~u_j~\mbox{is an arbitrary element such that}\\
  & \C{S}, [w \mapsto \Conc{u}] \models \nodeFormula{S}{u_j}(w)
        \end{array}
        \right .
\label{eq:CanonicCandidateEmbedding}
\end{equation}

Let us show that every concrete element is mapped to some element
in $U$. In the case that $Z(v_i)= \C{u}$, the concrete element $\C{u}$
is mapped to $u_i \in U$ by $\blur$.
Otherwise, because $\C{S} \models \characteristicFormula{S}[total]$
holds, at least one of its disjuncts must be satisfied by each $\C{u}$,
i.e., $\C{S}, [w \mapsto \C{u}]$ must satisfy $\nodeFormula{S}{u_i}(w)$ for
some $u_i$; thus $\blur$'s definition will map $\C{u}$ to this $u_i$.
Therefore, $\blur(\C{u})$ is well-defined.

In addition, every element $u_i \in U$ is assigned by $\blur$ to some concrete
element $\C{u}_i \in \C{U}$ such that $Z(v_i)= \C{u}_i$.
According to \lemref{unique}, all such elements $\C{u}_i$ are different.
%Since $\C{S}, \C{Z}$ satisfies the sub-formula $\Land_{k \neq j} \neg
%eq(v_k, v_j)$, all such elements $\C{u}_i$ are different.
Therefore, $\blur(\C{u})$ is surjective.

We shall show that $\blur$ satisfies \equref{tightEmbed} and \equref{canonicName};
that is, $\blur$ identifies $S$ as the canonical abstraction of $\C{S}$.

First, let us show that \equref{canonicName} holds for the abstraction imposed by $\blur$, namely that
a predicate $p$ in $S$ has the most precise abstract value w.r.t. the concrete values
that it represents, as is imposed by $\blur$.

Because $S$ is an \ICA, all nullary predicates in $S$ must have definite values, by \lemref{ICAProp}(ii).
$\C{S}$ satisfies $\characteristicFormula{S}_{nullary}$;
therefore, by \defref{pb}, nullary predicates in $\C{S}$ must have the same definite values as in $S$;
this shows that \equref{canonicName} holds for nullary predicates.

Because $S$ is an \ICA, all unary predicates in $S$ must have definite values, by \lemref{ICAProp}(iii).
Let $p$ be a unary predicate and let $u \in U$ be an individual of $S$ such that $\iota^S(p)(u) = b$.
We shall show that $p$ has the same definite value $b$ on
all concrete elements mapped to $u$ by $\blur$.
Because the join of these values is also $b$, we will get that \equref{canonicName} holds
for $p$ and $u$.
Recall that $\C{S}$ satisfies formula $\characteristicFormula{S}[p]$,
hence each assignment to $w$ satisfies the conjunct
$\nodeFormula{S}{u}(w) \implies p^b(w)$ of $\characteristicFormula{S}[p]$.
Let $\C{u} \in \C{U}$ be an individual of $\C{U}$ such that $\blur(\C{u})=u$
and consider an assignment in which $w$ is mapped to $\C{u}$.
By the definition of $\blur$, this assignment satisfies $\nodeFormula{S}{u}(w)$,
the premise of the conjunct. Therefore, it satisfies the conclusion, i.e.,
$\C{S}, [w \mapsto \C{u}]$ satisfies $p^b(w)$.
Using \defref{pb} we get that $\iota^{\C{S}}(p)(\C{u}) = b$.

Let $p$ be a predicate of arity $r > 1$.
If $p$ has a definite value $b$ in $S$ on a tuple $u_1, \ldots, u_r$,
$\characteristicFormula{S}[p]$
requires that $p$ evaluates to the same definite value $b$
on every concrete tuple $\C{u}_1, \ldots, \C{u}_r$ such that $\blur(\C{u}_i) = u_i$
(by the same argument as for unary predicates).
Therefore, the join operation returns $b$ as the most precise abstract value of $p$
for these concrete tuples.
Otherwise, if $p$ evaluates to $1/2$ on $u_1, \ldots, u_r \in U$,
there must be two tuples of elements in $\C{U}$,
say $\C{u}_{01}, \ldots, \C{u}_{0r}$ and $\C{u}_{11}, \ldots, \C{u}_{1r}$,
such that $\C{S}, [w_1 \mapsto \C{u}_{01}, \ldots, w_r \mapsto \C{u}_{0r}] \models \neg p(w_1, \ldots, w_r)$
and $\C{S}, [w_1 \mapsto \C{u}_{11}, \ldots, w_1 \mapsto \C{u}_{1r}] \models p(w_1, \ldots, w_r)$,
because $\C{S} \models \tau^S[p]$.
Thus, $p$ evaluates to $0$ on the first tuple and to $1$ on the second tuple
of the concrete structure;
therefore, the most precise value obtained by the join operation on these values is $1/2$.

We shall show that $\blur$ satisfies \equref{tightEmbed}, i.e., it
maps elements according to their canonical names. This involves showing two directions:
\begin{description}
\item [1.]For the sake of contradiction, assume that there are two distinct elements
$\C{u}_0, \C{u}_1 \in \C{U}$ that have the same canonical name (meaning that
for all $p \in \Voc_1$, $\iota^{\C{S}}(p)(\C{u}_0) = \iota^{\C{S}}(p)(\C{u}_1)$),
but $\blur(\C{u}_0) \neq \blur(\C{u}_1)$.
Because $S$ is a bounded structure,
there must be unary predicate $p$ that evaluates to $0$ on $\blur(\C{u}_0)$ and
to $1$ on $\blur(\C{u}_1)$.
As shown above, $p$ evaluates to the same definite values in the concrete structure $\C{S}$:
$\iota^{\C{S}}(p)(\C{u}_0) = 0$,
and $\iota^{\C{S}}(p)(\C{u}_1) = 1$ and a contradiction is obtained.
\item [2.] For the sake of contradiction, assume that two concrete elements,
denoted by $\C{u}_0, \C{u}_1 \in \C{U}$, have different canonical names,
but are mapped by $\blur$ to the same same element in $U$: $\blur(\C{u}_0) = \blur(\C{u}_1)$,
denoted by $u$.
By definition of $\blur$, $\C{S}, [w \mapsto \C{u}_i]$ satisfies
$\nodeFormula{S}{\blur(\C{u}_i)}(w)$, for $i=0,1$,
in other words $\C{S}, [w \mapsto \C{u}_i]$ satisfies $\nodeFormula{S}{u}(w)$.
Therefore, it satisfies each conjunct of $node$ formula,
i.e., for all $p$, $\C{S}, [w \mapsto \C{u}_i]$ satisfies
$p^{\iota^S(p)(u)}(w)$.
From this and the fact that all unary predicates in $S$ have
definite values because $S$ is an \ICA, we conclude by \defref{pb},
that $\iota^{\C{S}}(p)(\C{u}_i) = \iota^S(p)(u)$.
Therefore, $\iota^{\C{S}}(p)(\C{u}_0) = \iota^S(p)(u)$
and $\iota^{\C{S}}(p)(\C{u}_1) = \iota^S(p)(u)$,
for all $p \in \Voc_1$. Therefore, $\C{u}_0$ and $\C{u}_1$ have
the same canonical name and a contradiction is obtained.
\end{description}
\end{Proof}
\end{SubLemma}
%________________________________________________________________________
\begin{SubLemma}\label{Lem:CanonicRepresentOnlyIf}
For every $3$-valued structure $S$ that is an \ICA\,
and $2$-valued structure $\C{S}$ such that $\C{S} \models F$,
such that $S$ is the canonical abstraction of $\C{S}$, $\C{S} \models \tau^S$.
\begin{Proof}
Let $\blur \colon \C{U} \to U$ be the mapping that identifies $S$ as the canonical abstraction of $\C{S}$.
$\blur$ is a surjective function and possesses the properties in \equref{tightEmbed} and \equref{canonicName}.

First, we show that $\C{S} \models \characteristicFormula{S}$.
Let $\C{u}_i$ be an arbitrary element such that $\blur(\C{u}_i) = u_i$.
Define an assignment $\C{Z}$ such that $\C{Z}(v_i) = \C{u}_i$;
$\C{u}_i$ must exist because $\blur$ is surjective.
Because $S$ is canonical-FO-identifiable, by \lemref{CanonicBounded}
we conclude that for every $1 \leq i \leq n$,
$\C{S}, \C{Z} \models \nodeFormula{S}{u_i}(v_i)$.
According to \lemref{unique}, all the $\C{u}_i$ are distinct elements.
%Because $\blur$ is a function, all the $\C{u}_i$ are distinct elements,
%meaning that $\C{S}, \C{Z}$ satisfies the sub-formula $\Land_{k \neq j} \neg eq(v_k, v_j)$.

Because $\blur$ is a function, for every $\C{u}$ there is a $u$ such that $\blur(\C{u}) = u$.
Then, by \defref{CanonicUniqueFONames}, $\C{S}, [w \mapsto \C{u}] \models \nodeFormula{S}{u}(w)$,
i.e., every assignment to $w$ in $\C{S}$ satisfies some disjunct of $\characteristicFormula{S}_{total}$.
That is, $\C{S}$ satisfies $\characteristicFormula{S}_{total}$.

Because $S$ is an \ICA, nullary predicates have the same definite values in $S$ and in $\C{S}$, by \lemref{ICAProp}(ii).
Therefore, by \defref{pb},
$\C{S}$ satisfies $p^{\iota^{S}(p)()}$, for every nullary predicate $p \in \Voc_0$,
which means that $\C{S}$ satisfies $\characteristicFormula{S}_{nullary}$.

Let $p \in P$ be a predicate of arity $r$.
Let $\C{u}_1, \ldots, \C{u}_r \in \C{U}$ and let $\C{Z}$ be an assignment
such that $\C{Z}(w_i) = \C{u}_i$.
We shall show that $\C{S}, \C{Z}$ satisfies the body of \equref{naryFormula}.
Consider a conjunct of the body.
If the premise of the implication in this conjunct is not satisfied,
then the conjunct vacuously holds.
Otherwise, $\C{S}, \C{Z} \models \nodeFormula{S}{u_i}(w_i)$ for all $i = 1, \ldots, r$.
Then, by \lemref{CanonicBounded}, $\blur(\C{u}_i) = u_i$.
We have two cases to consider: (i)~if $\iota^S(p)(u_1, \ldots, u_r) = b \in \{1,0\}$ then
by \equref{canonicName} $\iota^{\C{S}}(p)(\C{u}_1, \ldots, \C{u}_r) = b$,
in other words, $\C{S},\C{Z}$ satisfies $p^b(w_1, \ldots, w_r)$.
(ii)~if $\iota^S(p)(u_1, \ldots, u_r) = 1/2$
then by \defref{pb}, $p^{\iota^S(p)(u_1, \ldots, u_r)}(w_1, \ldots, w_r) = p^{1/2}(w_1, \ldots, w_r) = \TRUE$,
which holds for any assignment.

To complete the proof, we show that for every $p \in \Voc_r$ of arity $r > 1$,
$\tau^S[p]$ holds.
Let $p$ be a predicate that evaluates to $1/2$ on a tuple $u_1, \ldots, u_r \in S$.
Because $S$ is an \ICA\,
$\iota^S(p)(u_1, \ldots, u_r) = 1/2$ means that the join operation in \equref{canonicName} yields $1/2$.
By the definition of join as the least upper bound, and using the
information order in \defref{ThreeValuedOrder},
we conclude that (i)~$\C{S}$ must contain at least two distinct tuples;
denoted by $\C{u}_{01}, \ldots, \C{u}_{0r}$ and $\C{u}_{11}, \ldots, \C{u}_{1r}$.
Because $\blur(\C{u}_{ij}) = u_j$ for $i=0,1$ and $j=1,\ldots,r$,
by \lemref{CanonicBounded} we get that
$\C{S}, [w \mapsto \C{u}_{ij}] \models \nodeFormula{S}{u_j}(w)$.
Therefore, each tuple satisfies $\Land_{j=1}^r \nodeFormula{S}{u_j}(w_j)$.
(ii)~$p$ evaluates to $0$ on the first tuple and $1$ on the second tuple.
This shows that $\C{S} \models \tau^S[p]$.
\end{Proof}
\end{SubLemma}
%________________________________________________________________________

\begin{SLem}{noIntersection}
Denote by $\domain$ the set of all $2$-valued structures
that satisfy the integrity formula $F$:
$\domain \eqdef \{ \C{S} \in \TSTRUCT{\Voc} \mid \C{S} \models F \}$.
Let $S$ be an ICA structure.
There exists a set of ICA structures $X$ such that
$\gamma_c(X) = \domain \smallsetminus \gamma_c(S)$.
\begin{Proof}
Denote by $Y$ the set of all ICA structures over a fixed vocabulary $\Voc$, i.e., $\gamma_c(Y) = \domain$.
We claim that $X$ is defined by $Y \smallsetminus S$.
By definition, $\gamma_c(X) = \gamma_c(Y \smallsetminus S)$,
and we show that $\gamma_c(Y \smallsetminus S) = \gamma_c(Y) \smallsetminus \gamma_c(S)$.
By the definitions of $Y$ and $\gamma_c$ in \equref{CanonicConcretization},
$\gamma_c(Y \smallsetminus S) \supseteq \domain \smallsetminus \gamma_c(S)$ holds.
To complete the proof, we show that the other direction of inclusion holds as well.
For the sake of argument, assume that there exists a $2$-valued structure $\C{S}$
that belongs to both $\gamma_c(S)$  and $\gamma_c(Y \smallsetminus S)$.
Thus, by \defref{ConcreteCanonic}, there exists an ICA structure $S'$ such that $\blur(\C{S}) = S'$,
and $S'$ is different from $S$. From \equref{canonicName}, it follows that
$\blur(\C{S}) \neq S$, which contradicts the assumption that $\C{S} \in \gamma_c(S)$.
\end{Proof}
\end{SLem}

%________________________________________________________________________
\begin{SLem}{closedUnderNegation}
Consider the formula $\tau^S$ from \equref{tightChareteristicFormula}, for some ICA structure $S$.
There exists a set of ICA structures $X$, such that
the formula $F \land \neg \tau^S$ is equivalent to the formula $\gammaHat_c(X)$.
%Let $\varphi$ be a formula such that $\varphi \eqdef \gammaHat_c(S)$
%for some ICA structure $S$.
%There exists a set of ICA structures $X$, such that
%the formula $\neg \varphi$ is equivalent to the formula $\gammaHat_c(X)$.
\begin{Proof}
Let $\domain$ be the set of all $2$-valued structures
that satisfy the integrity formula $F$.
Let $X$ be the set of ICA structures  that describes the complement of $\gamma_c(S)$,
as given by \lemref{noIntersection}.
Let $\C{S}$ be a $2$-valued structure such that
$\C{S} \in \gamma_c(X)$ if and only if $\C{S} \in \domain \smallsetminus \gamma_c(S)$.
The right-hand side simplifies to $\C{S} \in \domain$  and $\C{S} \notin \gamma_c(S)$.
Applying \theref{RepresentingTightEmbeddingByFormulae},
we get that
$\C{S} \models \gammaHat_c(X)$ if and only if $\C{S}$ satisfies $F$ but does not satisfy $\gammaHat_c(S)$.
Using \equref{gamma_c_X}, this is equivalent to $\C{S} \models F \land \neg \tau^S$.
\end{Proof}
\end{SLem}

%________________________________________________________________________
\begin{SThe}{RepresentingStructuresByNPFormulae}
For every $3$-valued structure $S$,
and a $2$-valued structure $\C{S}$:
\[
\C{S} \in \gamma(S)~ \mbox{iff}~\C{S} \models
\gammaHat_{NP}(S)
\]
\begin{Proof}
In \lemref{NPRepresentIf}, we show that the if-direction holds, i.e.,
every concrete structure satisfying the
NP-characteristic formula $\gammaHat_{NP}$ is indeed in $\gamma(S)$.
In \lemref{NPRepresentOnlyIf} we show the only-if part.
\end{Proof}
\end{SThe}

%________________________________________________________________________
\begin{SubLemma}\label{Lem:NPRepresentIf}
Let $S$ be a logical structure with set of individuals
$U=\{u_1, u_2, \ldots, u_n\}$.
Then, for all $\C{S}$ such that
$\C{S} \models \gammaHat_{NP}(S)$, $\C{S} \in \gamma(S)$.
\begin{Proof}
Let $\C{S}=\B{\C{U}, \C{\iota}}$ be a concrete structure such that $\C{S} \models
\widehat{\gamma}(S)$.
We shall construct a surjective function $f\colon \C{U} \to U$ such that
$\C{S} \sqsubseteq^f S$.
Let $\C{Z}$ be an assignment such that $\C{S}, \C{Z}
\models \varphi$ where $\varphi$ is the body of $\NPcharacteristicFormula{S}$
without the existential quantifiers on sets.
Let $\C{Z}(V_i) = U_i \subseteq \C{U}$.
Consider the following definition:
\begin{equation}\label{eq:NPCandidateEmbedding}
f(\C{u}) = \{u_i \mid \C{u} \in U_i \}
\end{equation}

$f(\C{u})$ is a set of size at most $1$ because the pair
$\C{S}, \C{Z}$ satisfies the sub-formula $\NPcharacteristicFormula{S}_{disjoint}$.
%of $\NPcharacteristicFormula{S}$.
This insures that the sets $U_1, \ldots, U_n$ are disjoint, i.e., each concrete element belongs to at most one
set. For simplicity, we say that $f(\C{u}) = u_i$, whenever $f(\C{u}) = \{u_i\}$.

We shall show that every concrete element is mapped by $f$ to some element in $U$.
Because $\C{S}, \C{Z}$ satisfies $\characteristicFormula{S}_{total}$,
we conclude that every concrete element satisfies the formula $\nodeFormula{S}{u_i}(w)$ for some
$u_i$.
Also, $\nodeFormula{S}{u_i}(w)$ given in \defref{NPCharacteristic} is a membership test in the set $V_i$;
therefore, every concrete element must be a member of some set $U_i$.
Thus, $\C{u}$ is mapped to $u_i \in U$, by the definition of $f$ in \equref{NPCandidateEmbedding}.
This shows that $f$ is well-defined.

Because $\C{S}, \C{Z}$ satisfies $\models \NPcharacteristicFormula{S}_{non\_empty}[i]$ for $i=1, \ldots, n$,
it must be that every $U_i$ contains at least one element, say
$\C{u}_i$, that is mapped to $u_i$ by $f$.
Because the sets are disjoint, all such elements $\C{u}_i$ are different.
Therefore, $f$ is surjective.

Let $p$ be a nullary predicate.
Because $\C{S}$ satisfies $\characteristicFormula{S}_{nullary}$,
it must satisfy each conjunct, in particular
$\C{S} \models p^{\iota^{S}(p)()}$.
Using \lemref{Trivial} we get that
$\iota^{\C{S}}(p)() \sqsubseteq \iota^{S}(p)()$.

Let $p \in P$ be a predicate of arity $r \geq 1$.
Let $\C{u}_1, \C{u}_2, \ldots, \C{u}_r \in \C{U}$ and let us show that
\begin{equation}\label{eq:naryProof2}
\iota^{\C{S}}(p)(\C{u}_1, \C{u}_2, \ldots, \C{u}_r) \sqsubseteq
\iota^S(p)(f(\C{u}_1), f(\C{u}_2), \ldots, f(\C{u}_r))
\end{equation}
Let $\C{Z}_1$ be an extension of assignment $\C{Z}$ such that $\C{Z}_1(w_i) = \C{u}_i$ for $i= 1, \ldots, r$.
Because $\C{S}, \C{Z} \models \characteristicFormula{S}[p]$,
we conclude that
$\C{S}, \C{Z}_1$ satisfies the body of \equref{naryFormula}.
Consider the conjunct of the body with premise
$\Land_{j=1}^r \nodeFormula{S}{f(\C{u}_j)}(w_j)$.
%If the premise does not hold, it imposes no limitations on $p$'s interpretation
%under $S$ and \equref{naryProof2} holds.
By definition of $f$, $\C{S}, w_j \mapsto \C{u}_j$ satisfies $\nodeFormula{S}{f(\C{u}_j)}(w_j)$
for all $j=1, \ldots, r$,
which means that the premise is satisfied by\ $\C{S}, \C{Z}_1$.
Therefore, the conclusion must hold:
%Otherwise, if the premise holds for $\C{S}, \C{Z}_1$, then the conclusion must hold:
\newline $\C{S}, \C{Z}_1 \models p^{\iota^S(p)(f(\C{u}_{1}), \ldots, f(\C{u}_{r}))}(w_1, \ldots, w_r))$
and the result follows from \lemref{Trivial}.
\end{Proof}
\end{SubLemma}
%________________________________________________________________________

\begin{SubLemma}\label{Lem:NPRepresentOnlyIf}
For every $3$-valued structure $S$,
and $2$-valued structure $\C{S}$ such that $\C{S} \models F$ and
$\C{S} \sqsubseteq S$, $\C{S} \models \characteristicFormula{S}$.
\begin{Proof}
Let $f\colon \C{S} \to S$ be a surjective function such that
$\C{S} \sqsubseteq^f S$.
Define an assignment $\C{Z}$ such that $\C{Z}(V_i) = U_i \subseteq \C{U}$
and $U_i = \{ \C{u}_i \mid f(\C{u}_i) = u_i \}$.

Because $f$ is a surjective function, there must exist at least one concrete
element that is mapped to $u_i$ by $f$.
This element belongs to the set $U_i$.
Therefore, $\C{S}, \C{Z} \models \Land_{i=1}^n \NPcharacteristicFormula{S}_{non\_empty}[i]$.

Because $f$ is a well-defined function, it maps each concrete element to exactly one
element $u_i \in U$, which induces the set $U_i$.
Therefore, a concrete element cannot belong to more than one set;
hence $\C{S}, \C{Z} \models \Land_{k \neq j} \NPcharacteristicFormula{S}_{disjoint}[k, j]$.

Because $f$ is a function, $f$ maps every concrete element to some element in $U$.
Therefore, every concrete element belongs to some set,
i.e., satisfies some disjunct of $\characteristicFormula{S}_{total}$.
That is $\C{S}, \C{Z} \models \characteristicFormula{S}_{total}$.

For every nullary predicate $p \in \Voc_0$, using \equref{EmbeddingCondition}
and \lemref{Trivial}, we conclude that $\C{S}, \C{Z}$ satisfies $p^{\iota^{S}(p)()}$.
Therefore, $\C{S}, \C{Z} \models \characteristicFormula{S}_{nullary}$.

Let $p \in P$ be a predicate of arity $r$.
Let $\C{u}_1, \ldots, \C{u}_r \in \C{U}$ and let $\C{Z}_1$ be an extension of assignment $\C{Z}$
such that $\C{Z}_1(w_i) = \C{u}_i$.
We shall show that $\C{S}, \C{Z}_1$ satisfy the body of \equref{naryFormula}.
If the premise of the implication is not satisfied, then the formula vacuously holds.
Otherwise, $\C{S}, \C{Z}_1 \models \nodeFormula{S}{u_i}(w_i)$ for all $i = 1, \ldots, r$.
Then, by \defref{NPCharacteristic},
$\C{u}_i$ belongs to the set $U_i$. The definition of $U_i$ implies that $f(\C{u}_i) = u_i$.
Using \equref{EmbeddingCondition}, we get
$\iota^{\C{S}}(p)(\C{u}_1, \ldots, \C{u}_r) \sqsubseteq \iota^S(p)(f(\C{u}_1), \ldots, f(\C{u}_r))$
which means $\iota^{\C{S}}(p)(\C{u}_1, \ldots, \C{u}_r) \sqsubseteq \iota^S(p)(u_1, \ldots, u_r)$.
By \lemref{Trivial} we conclude that
$\C{S},\C{Z}$ satisfies $p^{\iota^S(p)(u_1, \ldots, u_r)}(w_1, \ldots, w_r)$.
\end{Proof}
\end{SubLemma}
%________________________________________________________________________

\end{document}